\documentstyle[preprint,tighten,floats,aps,epsfig,amssymb]{revtex}

\newcommand{\eg}{e.g.}
\newcommand{\ie}{i.e.}
\newcommand{\etal}{{\it et al}.\ }
\newcommand{\als}{$\alpha_s$}
\newcommand{\pT}{$p_{\perp}$}
\newcommand{\ppbar}{$p\bar{p}$}
\newcommand{\ccbar}{$c\bar{c}$}
\newcommand{\bbbar}{$b\bar{b}$}
\newcommand{\Lepto}{\textsc{lepto}}
\newcommand{\Pythia}{\textsc{pythia}}
\newcommand{\Jetset}{\textsc{jetset}}
\newcommand{\Pompyt}{\textsc{pompyt}}

\newcommand{\Pma}{I\!\!P}

\begin{document}

\draft
\title{~\protect\\~\protect\\~\protect\\Soft Color Interactions and 
Diffractive Hard Scattering \protect \\ at the Fermilab Tevatron}

\author{R.~Enberg$^a$, G.~Ingelman$^{ab}$, N.~T\^\i mneanu$^a$}
\address{$^a$ High Energy Physics, Uppsala University, 
Box 535, S-751~21 Uppsala, Sweden\\
$^b$ Deutsches Elektronen-Synchrotron DESY, 
Notkestrasse 85, D-22603 Hamburg, Germany}

\maketitle

\vspace{-31ex}
\noindent DESY 01-076\\ 
\noindent June 2001
\vspace{+29ex}

\begin{abstract}
An improved understanding of nonperturbative QCD can be obtained by the recently
developed soft color interaction models. Their essence is the variation of
color string-field topologies, giving a unified description of final states in
high energy interactions, \eg, diffractive and nondiffractive events in $ep$
and \ppbar. Here we present a detailed study of such models 
(the soft color interaction model and the generalized area law model)
applied to \ppbar, considering also the general problem of the underlying
event including beam particle remnants. With models tuned to HERA $ep$ data,
we find a good description also of Tevatron data on production of $W ,$ beauty
and jets in diffractive events defined either by leading antiprotons or by one
or two rapidity gaps in the forward or backward regions. We also give
predictions for diffractive $J/\psi$ production where the soft exchange
mechanism produces both a gap and a color singlet $c\bar{c}$ state in the same
event. This soft color interaction approach is also compared with
Pomeron-based models for diffraction, and some possibilities to experimentally
discriminate between these different approaches are discussed. 
\end{abstract}


\section{Introduction}\label{sec-intro}
A major unsolved problem in particle physics is to understand strong
interaction processes with a small (`soft') momentum transfer. The most
striking illustration of this is the confinement of quarks and gluons in
hadrons and the related hadronization process giving the observable hadronic
final states in high energy collisions. In terms of Quantum Chromodynamics
(QCD), small momentum transfers have a large coupling $\alpha_s$ such that a
perturbative expansion in terms proportional to powers of $\alpha_s^n$ does
not work. This is in contrast to processes with a `hard' scale, \ie, a
large momentum transfer, where $\alpha_s$ is small and perturbative QCD (pQCD) 
on the level of quarks and gluons works well. To gain understanding of soft, 
nonperturbative QCD (non-pQCD) it is therefore advantageous to first 
consider soft effects in hard scattering events, since the hard scale gives a 
firm ground in terms of a parton level process which is calculable in pQCD. 
This hard-soft interplay is the basis for the topical research field of 
diffractive hard scattering \cite{StCroix}. 

Diffractive events can be characterized by having a rapidity gap, \ie, a region
in rapidity (or polar angle) without any particles. Another definition is to
require a leading particle carrying most of the beam particle momentum
($x_F\gtrsim 0.9$), which is kinematically related to a rapidity gap. These
rapidity gaps in the forward or backward rapidity regions, connect to the soft
part of the event and therefore nonperturbative effects on a long space-time
scale are certainly important. The central rapidity gaps between
high-$p_\perp$ jets, observed at the Tevatron \cite{j-g-j}, may be of a
different kind since the hard momentum transfer is across the gap. This gap
phenomenon will therefore not be considered here, but is studied separately
\cite{singlet-exchange}. 

Diffractive scattering has traditionally been explained in the Regge framework
by the exchange of a Pomeron \cite{GoulianosReview}. For processes
with a hard scale, a  parton structure of the Pomeron may be considered
\cite{IS}. With the Pomeron flux given by Regge phenomenology, the HERA data
on diffractive deep inelastic scattering can be well described by fitting
parton density functions in the Pomeron \cite{HERA-pomeron,HERA-F2D}. However, 
applying exactly the same model for $p\bar{p}$ gives a too large cross section 
for diffractive hard processes. Compared to the Tevatron data in
Table~\ref{tab-gapratios}, such a Pomeron model gives about a factor six too
large rates of $W$ and dijets with one gap and two orders of magnitude too
large rates of dijets with two gaps \cite{Alvero}. This is related to the
failure of the factorization theorem for hard diffractive hadron-hadron
scattering, although it holds in diffractive deep inelastic scattering (DIS)
\cite{Collins-Factorization}. It is also an indication of a
non-universality problem of the Pomeron model, which may be related to the
Pomeron flux. Since this flux specifies the leading particle spectrum, it is
interesting to note that the new Tevatron data~\cite{CDF-AP} with a leading
antiproton show a similar problem of the Pomeron model. These problems of the
Pomeron approach are further discussed in Section~\ref{sec-pomeron}. 

In order to better understand nonperturbative dynamics and to provide a unified 
description of all final states, the soft color interaction (SCI) model
\cite{SCI,unified} and the generalized area law (GAL) model \cite{GAL} were 
developed. These are added to Monte Carlo generators (\Lepto \cite{Lepto} 
for $ep$ and
\Pythia \cite{Pythia} for $p\bar{p}$) which simulate the interaction dynamics
and provide a complete final state of observable particles, such that an
experimental approach can be taken to classify events depending on the
characteristics of the final state: \eg, gaps or no-gaps, leading protons or
neutrons, etc. 

The basic assumption of the models is that variations in the topology of the
confining color  force fields (strings \cite{lund}) lead to different hadronic
final states after hadronization. 
The pQCD interaction gives a set of partons with a specific color order. 
However, this order may change owing to soft, nonperturbative
interactions. The details of our models for such interactions are described in
Section~\ref{sec-model}. One may at first think that this approach is some
kind of model for the Pomeron. To the extent that the term `Pomeron' is
associated with the Regge approach, this is not the case since nothing from
the Regge formalism is being used or referred to. The soft color interaction
models also give quite different results when applied to diffractive hard
scattering at the Tevatron. An overall summary of the relative rates of
various diffractive hard processes is given in Table~\ref{tab-gapratios},
which shows that this approach can account for several different gap phenomena
(taking the uncertainty in models and data into account). The details of this
and other results are presented and discussed in Sections~\ref{sec-single},
\ref{sec-dpe} and \ref{sec-jpsi}.

\begin{table}
\caption{Ratios diffractive/inclusive for hard scattering processes 
in \ppbar \ collisions at the Tevatron, showing experimental results
from CDF and D0 compared to the SCI and GAL soft color exchange models.}
\label{tab-gapratios} 
\begin{tabular}{lcllccc}
Observable & $\sqrt{s}$ &            & \multicolumn{3}{c}{Ratio [\%]} \\
           & [GeV]      & Experiment & Observed & SCI  & GAL  \\ 
\hline
$W$  - gap & 1800 & CDF \cite{CDF-W}    & $1.15 \pm 0.55$ & 1.2 & 0.8 \\

$Z$  - gap & 1800 &  --- &  ---\tablenotemark[1]  & 1.0 & 0.5 \\

$b\bar b$  - gap & 1800 & CDF \cite{CDF-B}    & $0.62 \pm 0.25$ & 0.7 & 1.4 \\

$J/\psi$  - gap & 1800 & ---    & ---\tablenotemark[1] & 1-2\tablenotemark[2] 
& 1-2\tablenotemark[2] \\

$jj$ - gap & 1800 &CDF \cite{CDF-JJ}   & $0.75 \pm 0.10$ & 0.7 & 0.6 \\ 

$jj$ - gap & 1800 &~~D\O \cite{D0-JJ}& $0.65 \pm 0.04$ & 0.7 & 0.6 \\ 

$jj$ - gap & ~~630  &~~D\O \cite{D0-JJ}& $1.19 \pm 0.08$ & 0.9 
& 1.2 \\ 

gap - $jj$ -gap\tablenotemark[3]& 1800 & CDF \cite{CDF-DP} 
& $ 0.26 \pm 0.06 $ & 0.2 &0.1\\

$\bar{p}$ - $jj$ -gap\tablenotemark[3] & 1800 &CDF \cite{CDF-DPE}
& $0.80 \pm 0.26$ &0.5& 0.4 
\end{tabular}

\tablenotemark[1]{~No result available}\\
\tablenotemark[2]{~Depending on kinematical requirements for $J/\psi$}\\
\tablenotemark[3]{~Ratio of two-gap events to one-gap events}\\

\end{table}

As opposed to the standard Pomeron approach, the SCI and GAL models can
describe diffractive events both at HERA and at the Tevatron. This is not
achieved by introducing several free parameters. On the contrary, the models
have essentially only {\it one} new parameter to account for the unknown
nonperturbative dynamics. This parameter is determined from the HERA data on
the diffractive structure function $F_2^D$ \cite{HERA-F2D} and then used with
the same computer code implemented in \Pythia \ to simulate \ppbar \ at the
Tevatron. 

The SCI and GAL models are very general in that they are able to describe a
large set of different data. This does not only refer to diffraction, but also
various nondiffractive observables. Particularly noteworthy is that the SCI
model reproduces the observed rate of high-$p_\perp$ charmonium and
bottomonium at the Tevatron \cite{SCI-onium}, which is factors of 10 larger
than the predictions based on the color singlet model in conventional pQCD.
Although the SCI and GAL models are too simple and have too weak theoretical
content to provide a satisfactory understanding, their general applicability
and success in describing different kinds of observables show that different
phenomena may have a common explanation. They represent a new approach which,
together with others mentioned in the Conclusions, may lead us towards a proper
understanding of nonperturbative QCD.

\section{Pomeron problems}\label{sec-pomeron}
The inability to describe both HERA and $p\bar{p}$ collider data on 
hard diffraction is a problem for the Pomeron model. It shows that 
the `standard' Pomeron flux factor \cite{DLpomeron},  
\begin{equation}\label{eq:pomeron-flux}
f_{\Pma /p}(x_{\Pma},t)=\frac{9\beta_0^2}{4\pi^2}
\left( \frac{1}{x_{\Pma}}\right) ^{2\alpha_{\Pma}(t)-1} 
\left[ F_1(t)\right]^2
\end{equation}
and Pomeron parton densities, $f_{i/\Pma}(x,Q^2)$, cannot be used 
universally. This flux is found to give a much larger cross section for
inclusive single
diffraction than measured at $p\bar{p}$ colliders, although it works well for
lower energy data. This is due to the increase of the flux as the minimum
$x_{\Pma\, min}=M^2_{X\, min}/s$ gets smaller with increasing cms energy
$\sqrt{s}$.
To avoid this unphysical increase, a Pomeron flux
`renormalization' has been proposed \cite{fluxrenorm} by enforcing that the
integral of the flux saturates at unity (by dividing by the integral whenever
it is larger than unity). This prescription not only gives the correct
inclusive single diffractive cross section at collider energies, but it also
makes the HERA and Tevatron data on hard diffraction compatible with the
Pomeron hard scattering model. The model result for HERA is not affected, but
at the higher energy of the Tevatron the Pomeron flux is reduced such that the
data are essentially reproduced. In another proposal \cite{Erhan-Schlein}
based on an analysis of single diffraction cross sections, the Pomeron flux is
reduced at small $x_{\Pma}$ through an $x_{\Pma}$- and $t$-dependent damping
factor. The pros and cons of these two approaches to modify the Pomeron flux
have been debated. 
 
It has recently been shown \cite{CFL-Wproduction} that the Tevatron data on
diffractive $W$ production can be reproduced if a harder Pomeron flux is
introduced together with a Pomeron intercept higher than the value extracted
from HERA data. These changes from the conventional Pomeron model illustrate
the problem of having a universally applicable Pomeron model. In a proposed 
new phenomenological approach \cite{Dino-new} the structure of the Pomeron 
is derived from that of the parent proton such that the gap probability is 
obtained from the soft parton density at $x_{\Pma}$. Some general features 
of diffractive DIS are obtained, but a more detailed confrontation with data 
remains to be performed. 

A difference between diffraction in $ep$ and $p\bar{p}$ is the possibility 
for coherent Pomeron interactions in the latter 
\cite{CollinsFrankfurtStrikman}.
In the incoherent interaction only one parton from the Pomeron participates
and any others are spectators. However, in the Pomeron-proton interaction 
with $\Pma =gg$ both gluons may take part in the hard interaction
giving a coherent interaction. For example, in the $\Pma p$ hard scattering 
subprocess $gg\to q\bar{q}$, the second gluon from the Pomeron 
may couple to the gluon from the proton. 
Such diagrams cancel when summing over all final states for the inclusive 
hard scattering cross section (the factorization theorem). 
For gap events, however, the sum is not over all final states and 
the cancellation fails leading to factorization breaking for these coherent
interactions where the whole Pomeron momentum goes into the hard scattering 
system. This coherent interaction cannot occur in the same way in deep 
inelastic
scattering (DIS) since the Pomeron interacts with a particle without colored
constituents. This difference between $ep$ and $p\bar{p}$ means that there
should be no complete universality of parton densities in the Pomeron. 
The difference between diffractive hard scattering at HERA and the Tevatron
can be described in terms of an overall suppression factor or gap `survival 
probability', due to extra soft rescattering effects in $p\bar{p}$, estimated 
using an eikonal model \cite{Khoze}. 

Although modified Pomeron models may describe the rapidity gap 
events reasonably well, there is no satisfactory understanding of the 
Pomeron and its interaction mechanisms. On the contrary, there are 
conceptual and theoretical problems with this framework. The Pomeron is 
not a real state, but only a virtual exchanged spacelike object. 
The concept of a
structure function is then not well defined and, in particular, it is unclear
whether a momentum sum rule should apply. In fact, the factorization into 
a Pomeron flux and a Pomeron structure function cannot be uniquely defined 
since only the product is an observable quantity \cite{Landshoff-Paris}. 

It may therefore be improper to regard the Pomeron as being `emitted' by the
proton, having QCD evolution as a separate entity and being `decoupled' from
the proton during and after the hard scattering. Since the Pomeron-proton
interaction is soft, its time scale is long compared to the short space-time
scale of the hard interaction. It is therefore natural to expect soft
interactions between the Pomeron system and the proton both before and after
the snapshot of the high-$Q^2$ probe provided by the hard scattering. The
Pomeron can then not be viewed as decoupled from the proton and, in
particular, is not a separate part of the QCD evolution in the proton. 

Large efforts have been made to understand the Pomeron as a two-gluon system
or a gluon ladder in pQCD. By going to the soft limit one may then hope to 
gain understanding of non-pQCD and, perhaps,  establish a connection 
between pQCD in the small-$x$ limit and Regge theory. 
More explicitly, diffractive DIS has been considered in terms of models 
based on two-gluon exchange in pQCD, see \eg \cite{pQCD-pomeron}. 
The basic idea is to take two gluons in a color singlet state from the 
proton and couple them to the $q\bar{q}$ system from the virtual photon. 
With higher orders included the diagrams and calculations become quite 
involved. Nevertheless, these approaches can be made to describe the main  
features of the diffractive DIS data. Although this illustrates the 
possibilities of the pQCD approach to the Pomeron, one is still forced to 
include nonperturbative modeling to connect the two gluons in a soft 
vertex to the proton which goes beyond the conventional use of parton
densities. Thus, even if one can gain understanding by working as far as
possible in pQCD, one cannot escape the fundamental problem of understanding
non-pQCD. 

\section{Models for soft color interactions}\label{sec-model}
Given these practical and conceptual problems of the Regge-based Pomeron model
and the impossibility to cover all important aspects by a pQCD treatment, new
approaches should be investigated. We are here exploring new ideas to model
non-pQCD interactions, which avoid the concept of a Pomeron and provide a
single simple model that describes all final states, with or without rapidity
gaps. 

The starting point is that the hadronic final state is produced through the 
hadronization of partons emerging from a hard scattering process which can 
be well described by pQCD. The basic new idea is that there
may be additional soft color interactions at a scale below the cutoff $Q_0^2$
for the perturbative treatment. Obviously, interactions will not disappear
below this cutoff. On the contrary, they will be abundant due to the large
coupling \als{} at small scales. The question is rather how to describe these
interactions properly. Here, we introduce soft color interactions which do not
change the dynamics of the hard scattering, but change the color topology of
the state such that another hadronic final state emerges after hadronization.
This topology can be described in terms of color triplet strings and the
standard Lund model \cite{lund} can be used for a well established treatment of
the hadronization of any given string configuration. We have tried two
different ways to model the soft exchange of color-anticolor representing
nonperturbative gluon exchange. The soft color interaction model is formulated
in a parton basis with color exchanges between the partons emerging from the
hard scattering process (including remnants of initial hadrons). The
generalized area law model is instead formulated in a string basis, since
strings are here assumed to be the proper states for soft exchanges that may
not resolve partons. In spite of this difference, the models have a very
similar structure and may be regarded as variations on the same general theme.

The SCI and GAL models are constructed as subroutines added to the Monte Carlo
event generators \Lepto \cite{Lepto} for $ep$ and \Pythia \cite{Pythia} for
$p\bar{p}$. This gives powerful tools for detailed investigations of the
models and their ability to reproduce experimental data. 

Since the soft non-pQCD processes cannot alter the hard perturbative
scattering
processes, the latter should be kept unchanged in the models. Therefore, the
hard parton level interactions are treated in the normal way using standard
hard scattering matrix elements (electroweak or QCD) plus initial and final
state parton showers based on the DGLAP leading logarithm evolution equations
\cite{DGLAP} to simulate higher order pQCD processes. Thus, the set of
partons, including those in beam hadron remnants, are generated as in
conventional $ep$ and \ppbar \ hard scattering processes. The SCI and GAL
models are then added as an extra intermediate step before the hadronization
is performed using the Lund Monte Carlo {\sc jetset} \cite{Pythia}. 

In this section we first describe these two models in some detail and then
discuss other aspects of soft interactions which are common for both models
and must be considered in a complete Monte Carlo model. 

\subsection{The SCI model}\label{sec-SCI}
The soft color interaction (SCI) model \cite{SCI,unified} 
is applied to the parton state emerging from the hard scattering. It gives 
the possibility for each pair of these color charged partons to make a soft
interaction. One may here include all possible pairs of partons or require
that one parton belongs to the remnant. In the latter case, one may view this
as the perturbatively produced quarks and gluons interacting softly with the
color medium of the proton as they propagate through it. The soft interaction
changes  only the color but not the momentum and may be viewed as soft
nonperturbative gluon exchange.  This should be a natural part of the process
in which bare perturbative partons are dressed into nonperturbative ones
and the formation of the confining color flux tube in between them. This
necessarily involves some, not yet understood nonperturbative interactions
which the model attempts to describe. 

\begin{figure}[t]
\begin{center}
\epsfig{file=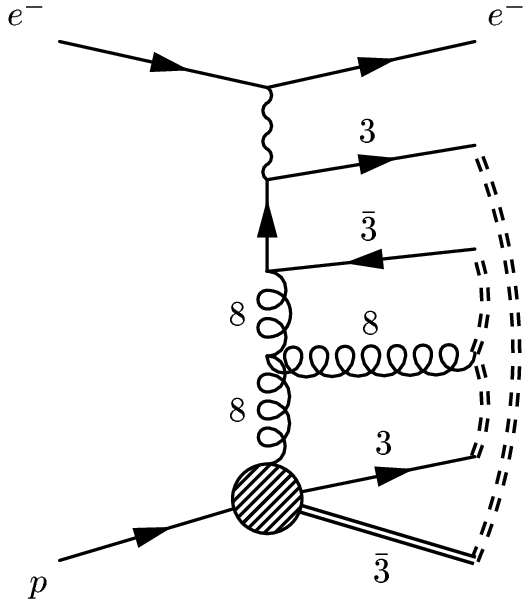, clip=, bbllx=23, bblly=619, bburx=160, 
bbury=782}
\epsfig{file=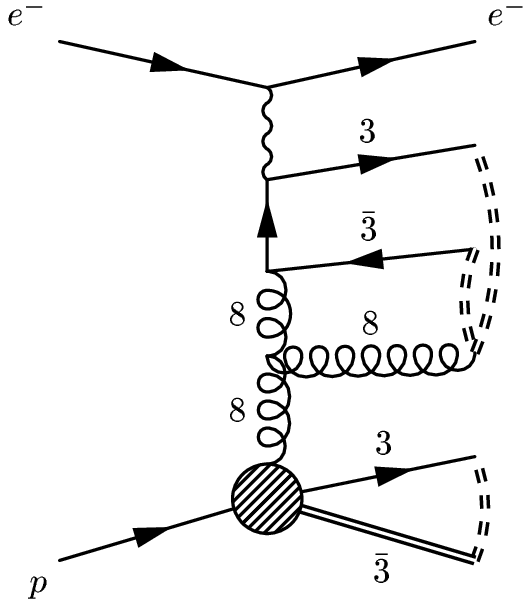, clip=, bbllx=23, bblly=619, bburx=160, 
bbury=782}
\vspace*{5mm}
\caption{Boson-gluon fusion process in DIS with string configuration 
(double-dashed line) in (left) the conventional Lund string model connection 
of partons and (right) after a soft color octet exchange between the remnant
and the hard scattering system as modeled by SCI or GAL resulting in a phase
space region without a string leading to a rapidity gap event after
hadronization.} \label{DIS-SCI}
\end{center}
\end{figure}

Being a nonperturbative  process, the exchange probability cannot presently be 
calculated and is therefore described by a phenomenological parameter $P$. The
number of soft exchanges will vary event-by-event and change the color
topology such that, in some cases, color singlet subsystems arise separated in
rapidity,  as illustrated Fig.~\ref{DIS-SCI} where, \eg, a color exchange
between the perturbatively produced quark and the quark in the remnant has
taken place. Color exchanges between the perturbatively produced partons and
the partons in the proton remnant (representing the color field of the proton)
are of particular importance for the gap formation. It should be emphasized,
however, that the model is quite general giving rise to events both with and
without rapidity gaps. 

Since DIS is a simpler and cleaner process than \ppbar \ collisions, the model
was first developed for DIS and successfully tested against diffractive DIS
data from HERA \cite{SCI,unified,sce-heramc}. The rate and main properties of 
the gap events are qualitatively reproduced. The rate of gap events depends on 
the parameter $P$, but the dependence is not strong giving a stable model with
$P\simeq 0.2$--0.5. This color exchange probability is the only new parameter
in the model. Other parameters belong to the conventional \Lepto \ model
\cite{Lepto} and have their usual values. The rate and size of gaps do,
however, depend on the amount of parton emission. In particular, more initial
state parton shower emissions will tend to populate the forward rapidity
region and prevent gap formation \cite{unified}.  

The gap events show the properties characteristic of diffraction. The 
exponential
$t$-dependence arises in the model from the intrinsic transverse momentum
(Fermi motion) of the interacting parton which is balanced by the proton
remnant system. This remnant gives rise to leading protons with a peak
at large fractional momentum $x_F$, as well as proton dissociation. 

The salient features of the measured diffractive structure function are also
reproduced \cite{sce-heramc}. The behavior of the data on $F_2^D(\beta,Q^2)$ is
in the SCI model understood as normal pQCD evolution in the proton. The rise
with $\ln{Q^2}$ also at larger $\beta$ is simply the normal behavior
at the small momentum fraction $x=\beta x_{\Pma}$ of the 
parton in the proton. Here, 
$x_{\Pma} =\frac{Q^2+M_X^2-t}{Q^2+W^2-m_p^2} 
\approx \frac{x(Q^2+M_X^2)}{Q^2}$ 
is only an extra variable related to the gap size or $M_X$, which does not
require a Pomeron interpretation. The flat $\beta$-dependence of 
$x_{\Pma} F_2^D=\frac{x}{\beta} F_2^D$ is due to the factor $x$ compensating
the well-known increase at small-$x$ of the proton structure function $F_2$.
 
This Monte Carlo model gives a general description of DIS, with and without 
gaps. In fact, it can give a fair account of such `orthogonal' observables 
as rapidity gaps and the large forward $E_T$ flow \cite{unified}. 
Diffractive events are in this model 
defined through the topology of the final 
state, in terms of rapidity gaps or leading protons just as in experiments.  
There is no particular theoretical mechanism or description in a separate 
model, like Pomeron exchange, that defines what is labeled as diffraction. 
This provides a smooth transition between diffractive gap events and 
nondiffractive (no-gap) events \cite{SCI-forward}. 
In addition, leading neutrons are also obtained in fair agreement with 
recent experimental measurements \cite{leading-pn}. In a conventional 
Regge-based approach, Pomeron exchange would be used to get diffraction, 
pion exchange added to get leading neutrons and still other exchanges added 
to get a smooth transition to normal DIS. The SCI model demonstrates that a 
simpler theoretical description can be obtained.  

\subsection{The GAL model}\label{sec-GAL}
The generalized area law (GAL) model \cite{GAL} for color string
re-interactions is a model for soft color exchanges which is similar in
spirit to the SCI model. Whereas the SCI model is 
formulated as soft exchanges
between the partons emerging from the hard scattering process, the GAL model
is formulated in terms of interactions between the strings connecting these
partons. Soft color exchanges between strings change the color topology
resulting in another string configuration, as illustrated in
Fig.~\ref{DIS-SCI}. 

The probability for two strings to interact is in the GAL model obtained as a
generalization of the area law suppression $e^{-bA}$ with the area $A$ swept
out by a string in energy-momentum space. The model uses the measure 
$A_{ij}=(p_i+p_j)^2-(m_i+m_j)^2$ for
the piece of string between two partons $i$ and $j$. This results in the
probability $P=P_0[1-\exp{(-b\Delta A)}]$ depending on the change $\Delta A$
of  the areas spanned by the strings in the two alternative configurations of
the strings, \ie, with or without the topology-changing soft color exchange.
The exponential factor favors making `shorter' strings, \eg, events with gaps,
whereas making `longer' strings is suppressed. The fixed probability for soft
color exchange in SCI is thus in GAL replaced by a dynamically varying one. 

There is only one new parameter in the GAL model, \ie, $P_0$ instead of $P$ in
SCI. $b$ is one of the usual hadronization parameters in the Lund model
\cite{lund}, but its value must be retuned when changing the string
configuration. 
Since the GAL model is formulated in terms of strings, it should be applicable
to all interactions producing strings, \ie, also to hadronic final states in
$e^+e^-$. 
The parameter values used in the GAL model were obtained
\cite{GAL} by making a simultaneous tuning to the diffractive structure
function in DIS and the charged particle multiplicity distribution and
momentum distribution for $\pi^{\pm}$ in $e^+e^-$ annihilation at the
$Z^0$-resonance. This resulted in $P_0=0.1$, $b=0.45$ GeV$^{-2}$ and $Q_0=2$
GeV, where $Q_0$ is the cutoff for initial and final state parton showers. It
is not possible to have the {\Jetset} default cutoff $Q_0=1$ GeV in the
parton showers and simultaneously reproduce the multiplicity distribution. One
might worry that the obtained cutoff is relatively large compared to the
default value. However, it is not obvious that perturbation theory should be
valid for so small scales when more exclusive final states are considered.
Therefore, $Q_0$ can be seen as as a free parameter describing the boundary
below which it is more fruitful to describe the fragmentation process in terms
of strings instead of perturbative partons.

With this parameter tuning the GAL model gives very similar results \cite{GAL}
for the final state in $e^+e^-\to Z^0\to hadrons$ as default \Jetset. This
concerns multiplicity distributions, momentum distributions and string
effects. Also the conventional rapidity gap behavior is obtained, \ie, an
exponentially falling distribution with increasing size $\Delta y$ of the
largest rapidity gap in the event. 

Applying the GAL model to DIS at HERA \cite{sce-heramc} gives a quite good
description of the diffractive structure function
$F_2^{D(3)}(x_{\Pma},\beta,Q^2)$ observed by H1. The details at low $Q^2$ is
actually better reproduced with GAL than with SCI. The GAL model cures the
problem the SCI model has in producing somewhat too many soft hadrons in 
inclusive DIS, but results in too low transverse energy flow in the forward 
region. These effects are related to events where the string after SCI goes 
back-and-forth producing a zig-zag shape, \ie, a longer string, giving more 
but softer hadrons after hadronization. Conversely, the GAL model suppresses 
topologies with long strings. 

\subsection{Remnants and soft underlying event}\label{sec-remnant}
To obtain a complete model for the production of the observable hadronic final
state there are further issues of nonperturbative dynamics that have to be
considered. These include not only the hadronization process itself, but also
the treatment of remnants of the colliding hadrons and possible additional
dynamics in order to achieve a decent description of the soft underlying
event, \ie, underlying the hard scattering part of the event. Here, we
essentially use the standard models developed for the family of Lund Monte
Carlo programs, but with some modifications and further developments as will
be described in this subsection. 

The standard Lund hadronization model \cite{lund} as implemented in \Jetset
\cite{Pythia} is used for the formation of hadrons from color triplet string
fields. However, the final state will depend on how the strings are stretched
between partons, as exemplified by the SCI and GAL models above. Similarly,
the resulting string system will depend on how the hadron remnants are
treated and if additional strings are formed, \eg, to produce additional
hadronic activity in the underlying event. 

The remnant system is the initial (anti)proton `minus' the parton entering the
hard scattering process, \ie, the hard $2\to 2$ scattering given by matrix
elements combined with parton showers. The initial parton carries a momentum
fraction $x_0$ of the beam proton as given by the parton density
distributions $f_i(x_0,Q^2_0)$ at the scale $Q_0^2$ where the initial state
parton shower is terminated in its backwards evolution simulation. This
leaves the fraction $1-x_0$ for the proton remnant system. The initial parton
can be either a valence quark, a sea quark or a gluon. In case a valence
quark is removed from the initial proton, the remnant is a diquark with an
anti-triplet color charge that defines the endpoint of a triplet
string. If the initial parton is a gluon, the remnant contains all three
valence quarks in a color octet state which is split into a color triplet
quark and a color anti-triplet diquark that form the end-points on two
triplet strings. Here, the quark and diquark share the remnant momentum in
the fractions $\chi$ and $1-\chi$, respectively, as given by parametrizations
of $\cal P(\chi)$ in \Pythia \ and to be further discussed below. 

In case a sea quark is removed from the initial proton, the remnant system is
more complex, containing all three valence quarks plus the partner of the
interacting sea quark in order to conserve quantum numbers. Here, a more
elaborate sea quark treatment (SQT) has been introduced \cite{Lepto,unified}.
The interacting quark, with flavor and momentum $x_0$ obtained from the initial
state parton shower evolution, is taken as a valence or sea quark based on
the relative sizes of the corresponding parton distributions
$q_{val}(x_0,Q_0^2)$ and $q_{sea}(x_0,Q_0^2)$. In case of a sea quark, the
left-over partner is given an explicit momentum. Here, we have tried two
possibilities to model this unknown dynamics. In the first (SQT1), the
longitudinal momentum fraction is given by the Altarelli-Parisi 
splitting function $P(g\to q\bar{q})$, \ie, the pQCD initial state parton shower
routine is used to model
a $g\to q\bar{q}$ process which is strictly speaking below the original
parton shower cutoff. As an alternative (SQT2) the sea quark partner is
assigned a longitudinal momentum chosen from the corresponding sea quark
momentum distribution in the proton. In both cases the transverse momentum 
is chosen from the same  
Gaussian used for the primordial transverse (Fermi) momentum. These two 
methods give similar results, but differ in some details as will be 
discussed below. The
sea quark partner and the three valence quarks, which are split into a quark
and a diquark as described, define the dynamics of the remnant system.
These three color (anti)triplet objects in the remnants are then end-points
on strings, implying additional string topology possibilities. Since the sea
quark partner has only a small transverse momentum, it affects in particular
the very forward part of the final hadronic state. Therefore, it is of
interest for the formation of rapidity gaps studied here. 

A related issue is the treatment of a color singlet system (string) with small
invariant mass. The Lund hadronization model is constructed for large mass
strings, but can be applied to systems of invariant mass which is as low as
the sum of the end-point parton masses plus an additional $\sim 1$ GeV. When 
the string mass is so small that only one or two hadrons can be formed, normal
string hadronization is not applicable since energy-momentum constraints and
resonance phenomena demand special treatment. This is instead achieved through
the new routines ({\sc lsmall} in  \Lepto \ and {\sc pysmall} in  \Pythia ). Of
particular importance for the investigations in this paper is the formation of
a single leading proton (or antiproton) giving the diffractive signature. The
mapping of a string with a continuous mass distribution onto a particular
on-shell hadron with fixed mass, requires a shuffling of energy-momentum to
another string system in order to conserve energy-momentum in the event
\cite{Pythia}. By transferring the required energy-momentum to/from another
parton which is as far away as possible in phase space, the relative
disturbance on the four-vectors is kept minimal and typically of order tens of
MeV, \ie, small even on the hadronization momentum transfer scale. 

Starting with the hard scattering processes 
(matrix elements and parton showers)
and adding this remnant treatment followed by Lund string model hadronization
results in a Monte Carlo event generator producing a complete hadronic final
state. The resulting hadronic activity is, however, too small compared to
collider data \cite{Multiple}. The observed multiplicities are larger, with the
multiplicity distribution extending in a longer tail to large multiplicities.
Furthermore, the number of particles per unit rapidity is larger and gives a
higher rapidity plateau or `pedestal' below high-\pT{} jets than obtained in
the model. This additional activity in the underlying event is related to soft
QCD processes and is therefore difficult to describe in a theoretically
satisfactory way. 

In \Pythia \ this additional activity in the underlying event is achieved by a
model for multiple interactions (MI) \cite{Multiple,Pythia}. This is
constructed based on multiple parton-parton scatterings described by the QCD
$2\to 2$ matrix elements. At small momentum transfers this cross section
becomes large, even larger than the \ppbar{} total cross section which is
interpreted as having more than one such parton-parton scattering in the same
event. These scatterings can sometimes be hard enough to contribute to the rate
of low-\pT{} jets and minijets, but dominantly they have too small \pT{} to
give observable jet structure. These small-\pT{} partons will stretch
additional strings that produce more hadrons over large rapidity regions and
thereby contribute substantially to the underlying event. 

The cross section for these multiple scatterings diverge when the scattered 
parton $p_\perp \to 0$. This is avoided by some (arbitrary) regularization or a
cutoff on \pT, which will be the main regulator of the amount of multiple
scatterings that are generated. In the default version of the MI model in 
\Pythia~5.7 a sharp cutoff $p_{\perp}^{\mathrm{min}}=1.4$ GeV is used, although
more complicated alternatives are available as options \cite{Pythia}. 
Using this MI model, data on multiplicities, rapidity distributions and pedestal
effects at the S\ppbar S ($\sqrt{s}=540$ and $630$ GeV) \cite{Underlying} can
be reasonably described \cite{Multiple}. Measurements of this kind have only
recently been made at the Tevatron and the model has not yet been tested or
tuned at the energy of interest in our study. 

Although the MI model is based on pQCD parton-parton scattering, in this context
the model is used to emulate soft nonperturbative effects. The soft color
exchange models are also introduced to account for soft effects on the hadronic
final state. The SCI model, in particular, can give zig-zag shaped strings
which
produce a larger number of hadrons per unit rapidity, \ie, more activity in the
underlying event. There is therefore a risk of `double counting' the soft
effects and producing too much underlying soft activity if the SCI/GAL model and
the MI model are simply added. With the SCI/GAL model tuned to data on rapidity
gaps, we therefore lower the amount of multiple interactions by increasing the 
$p_\perp^{\mathrm{min}}$ parameter. This means that the pQCD-based MI model is
not pushed to generate the softest dynamics, which is instead treated by the
soft exchange models. We have studied this issue in some detail by looking at
jet profiles, rapidity plateaus and charged particle multiplicities obtained by
running \Pythia{} with SCI/GAL added and the default MI model. Keeping the
default value of $p_\perp^{\mathrm{min}}$ gives, as expected, too much
underlying event activity, whereas increasing to $p_\perp^{\mathrm{min}}=2.5$
GeV for SCI and to $p_\perp^{\mathrm{min}}=2.0$ GeV for GAL, one obtains
essentially the same results as default \Pythia{}, and thereby reproduce data
equally well. The lower value for the GAL model reflects the fact that longer
strings are suppressed, and therefore GAL contributes less to the underlying
event activity than SCI. We note that in the recently released version 6 of
\Pythia{} \cite{Pythia6}, the MI cutoff has been made energy dependent giving
the value $p_\perp^{\mathrm{min}}=2.1$ GeV at the Tevatron, \ie, closer to our
values and indicating that the GAL model adds very little activity to the
underlying event. Our $p_\perp^{\mathrm{min}}$ values have also been obtained by
comparing with the diffractive data studied in this paper, but this will be
discussed further below. 

The sensitivity of our results to variations in these details of the modeling
of
the remnant and the underlying event has been investigated and is discussed
below in connection with the comparison of our models and the available data.

\section{Single diffractive hard scattering}\label{sec-single}

Before discussing the details of how the SCI and GAL models apply in different
single diffractive hard scattering processes in the following subsections, we
first discuss some general aspects. 

\begin{figure}[t]
\begin{center}
\epsfig{file=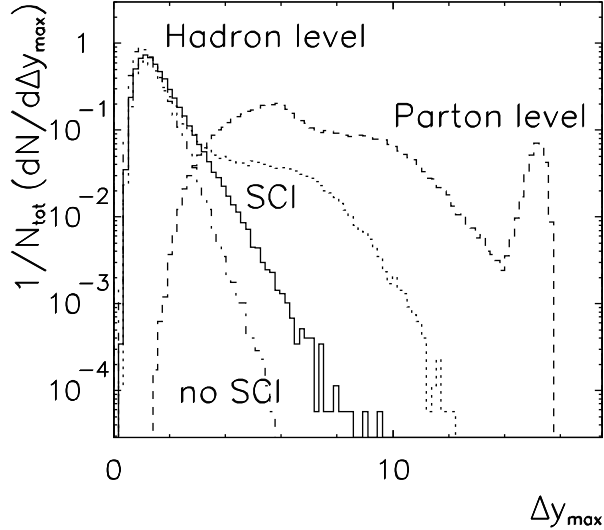} \\ 
\caption{Distribution of the size $\Delta y_{max}$ of the largest rapidity gap
in $W$ production in \ppbar \ events at $\sqrt{s}=1.8$ TeV in \Pythia . The
dashed curve represents the parton level obtained from hard, perturbative
processes. The dash-dotted curve is for the hadronic final state after standard
hadronization, whereas adding the soft color interaction model results in the
full curve and the dotted curve when setting $m_W=8$ GeV to show the appearance
of a `diffractive plateau' when the kinematical constraint of the $W$ mass is
relaxed. } \label{plotmaxgapsize}
\end{center}
\end{figure}

Single diffractive scattering is characterized by a large rapidity gap in the
forward or backward hemisphere of a \ppbar \ collision. The occurrence of
rapidity gaps is very strongly affected by soft effects, as demonstrated in
Fig.~\ref{plotmaxgapsize} for the case of diffractive $W$ production. At the
parton level, arising from the hard processes described by matrix elements and
parton showers, there can be large regions of phase space where no partons have
been emitted and thereby no strong suppression of the probability for large
rapidity gaps. The partons are, however, connected by color force fields which
through hadronization produce hadrons which fill these gaps in the final state.
Thus, applying hadronization using the standard Lund string model, causes the
drastic transition from the dashed  to the dash-dotted curve in
Fig.~\ref{plotmaxgapsize} such that large rapidity gaps in the final state of
hadrons become exponentially suppressed. An extreme case is provided by the
peak in the parton level curve, which arises from events where the $W$ is
produced by valence quark annihilation without parton radiation resulting in a
huge rapidity separation between the two remnant systems (diquarks).
Hadronization of the color string between these remnants produce hadrons in the
full rapidity range, leaving no trace of the parton level gap. 

This very strong effect of hadronization implies that modifications of the
modeling of the poorly known nonperturbative QCD processes can have substantial
effects. Applying the SCI model of last section, leads to an increased
probability for large rapidity gaps (full curve in Fig.~\ref{plotmaxgapsize})
at the hadron level, but still far below the parton level result. This
difference relative to default \Pythia \ may at first seem small, but for large
gaps it is exactly what is needed to describe data as will be discussed in
detail below. One may worry that there is no flat region, \ie, where the
probability does not decrease with increasing gap size, which is sometimes
taken as a characteristic for diffraction. This is due to the kinematical
restriction on high-$x_F$ leading protons imposed by the large $W$ mass, as 
verified in the Monte Carlo by lowering $m_W$ resulting in the expected
diffractive behavior shown by the dotted curve in Fig.~\ref{plotmaxgapsize}. 

\begin{figure}[t]
\begin{center}
\epsfig{width= 1.0\columnwidth,file=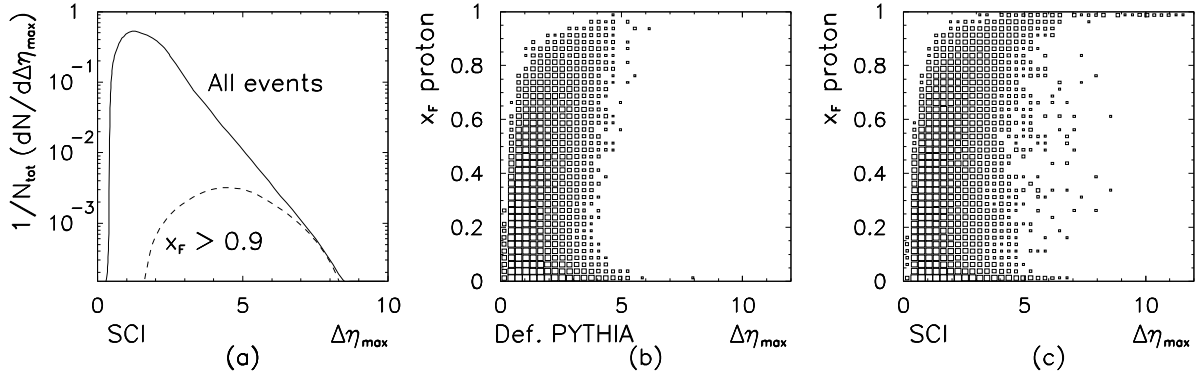}
\caption{Illustrations of the relation between the size $\Delta \eta_{max}$
of the largest rapidity gap and the momentum fraction $x_F$ of the leading
proton in simulated \ppbar \ events at $\sqrt{s}=1.8$ TeV. 
(a) Distribution of largest gap for all $W$ events (full curve) in the SCI model
and for the subsample having a leading proton with $x_F > 0.9$ (dashed).
(b,c) Scatterplot (logarithmic scale) showing the correlation between leading
proton $x_F$ and associated largest gap for dijet production in 
(b) default \Pythia\ and (c) when including the SCI model.}
\label{gapsize}
\end{center}
\end{figure}

Diffractive events can be defined experimentally in two different ways: by a
rapidity gap or by a leading (anti)proton. (Given the symmetry between proton
and antiproton beams at the Tevatron, we usually mean either proton or
antiproton when speaking of a leading proton.) The two methods are related,
since kinematics requires an event with a leading proton to also have a gap.
This has been explicitly investigated with our Monte Carlo model resulting in
Fig.~\ref{gapsize}. Events with a very large gap do typically have a leading
proton. At Tevatron energies, however, gaps of substantial size are
kinematically allowed also for protons with not so high $x_F$ as shown by the
nontrivial correlations in Fig.~\ref{gapsize}b and c. This calls for some
caution when comparing results based on these two definitions of diffractive
events. 
Irrespectively of this warning, comparing Fig.~\ref{gapsize}b and c shows the
effect of the SCI model to produce more events with large-$x_F$ protons and 
large gaps. When the leading proton is at a low $x_F$ there may be another 
leading system of small invariant mass, in particular a large-$x_F$ neutron. 
In addition to the events included in Fig.~\ref{gapsize}b and c, there is a 
substantial amount of gap events without a proton, but with other leading 
particles. Such events, which are natural products of the Monte Carlo model, 
must be included when using a gap definition of diffraction. The diffractive 
rates obtained with a gap definition are therefore usually larger than those 
obtained with a leading proton definition.  

The experimental results on diffractive hard scattering processes have mainly
been presented as relative rates, \ie, the cross section for a diffractive
process divided by the total cross section for the same hard process. We denote
this diffractive ratio by $R_{\mathrm{hard}}$, where `hard' stands for the
relevant hard subprocess. The first experimental analyses, \eg
\cite{CDF-W,CDF-B,CDF-JJ}, used a gap definition of diffraction. This is,
however, essentially equivalent to requiring a leading proton with $x_F>0.9$,
such that the diffractive ratio $R_{\mathrm{hard}}$
can be expressed as 
\begin{equation}
R_{\mathrm{hard}} = \frac{1}{\sigma_{\mathrm{hard}}^{\mathrm{tot}}}
\int_{{x_F}_{\mathrm{min}}}^1 dx_F \, \frac{d\sigma_{\mathrm{hard}}}{dx_F}. 
\label{RW}
\end{equation}
where ${x_F}_{\mathrm{min}}$ is the minimum leading proton $x_F$ for an event to
be considered diffractive. The values of $R_{\mathrm{hard}}$ in
Table~\ref{tab-gapratios} were obtained with ${x_F}_{\mathrm{min}}=0.9$,
corresponding to the experimental analyses. This is also in accordance with the
conventional definition of diffraction in the Regge approach and comparisons
with simulations of Pomeron exchange at $x_{\Pma}=1-x_F < 0.1$ were made,
leading to the problems discussed in Section \ref{sec-pomeron}. The variation
of $R_{\mathrm{hard}}$ with ${x_F}_{\mathrm{min}}$ in the models will be
discussed below. 

Some more recent CDF analyses \cite{CDF-AP,CDF-DPE} could define diffraction
through leading antiprotons observed in Roman pot detectors. This provided
additional information, on $x_{\Pma}$ and the momentum fraction $x$ of the
struck parton in the incoming antiproton, making the results less inclusive.
This gives additional handles to test the models, as will be discussed below.

Our results presented below were obtained by Monte Carlo simulations using
\Pythia{} version 5.7. As a reference, called `default', we use standard
\Pythia{} with all parameters and switches at their default values. The parton
distributions CTEQ3L~\cite{CTEQ3} were used for the simulations with default
\Pythia{} and with the SCI model, and CTEQ4L~\cite{CTEQ4} were used with GAL.
There is a slight variation of the results depending on this choice, see
Table~\ref{tab-RW} and the discussion in section \ref{Wprod}. The SCI and
GAL models are simulated using added subroutines as described in Section
\ref{sec-model}. This includes the improved procedures for beam particle
remnants, with the treatment of sea quark interactions and small mass string
systems. 

In order to compare with the Pomeron model, we have also included results from
simulations using the \Pompyt{} program (version 2.6) \cite{Pompyt}. Here, the
Donnachie-Landshoff (DL) \cite{DLpomeron} Pomeron flux and the
Gehrmann-Stirling (GS) \cite{GSpomeron} Pomeron structure functions were used. 
The GS parametrizations have two variants, referred to as model I and II. In
short, model I describes the Pomeron as a hadronic system of quarks and gluons.
 Model II has, apart from this `resolved' component, also a `direct' component
with a photon-Pomeron coupling. Both models have been tuned to describe HERA
data. We have mainly used model I, as this describes the Tevatron data better,
but we have also tested model II. Still other parametrizations of the Pomeron
structure function are available, but using them will not change the results in
an essential way. 

After having defined our models and described the general framework, we can now
turn to the specific diffractive hard scattering processes. 

\subsection{Diffractive W production}\label{Wprod}

\begin{figure}[t]
\begin{center}
\epsfig{file=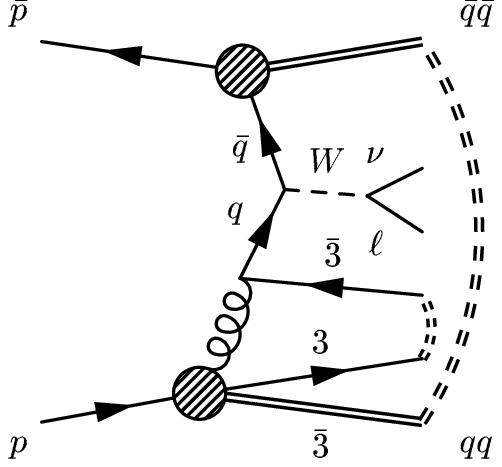}
\epsfig{file=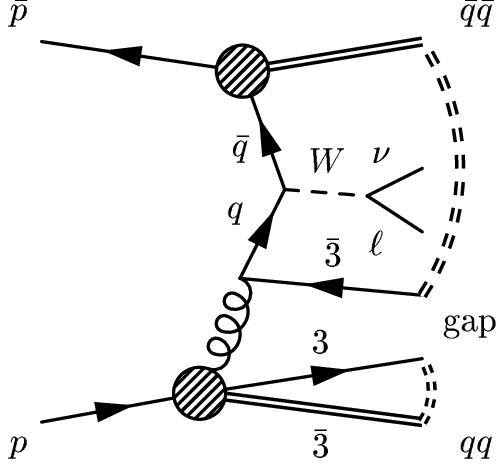}
\vspace*{5mm}
\caption{Production of $W$ (decaying leptonically) in a $p\bar p$ collision 
with string topology (double-dashed lines) before (left) and after (right) 
a soft color exchange resulting in a rapidity gap.} \label{fig:Wsci}
\end{center}
\end{figure}
\begin{figure}[t]
\begin{center}
\epsfig{file=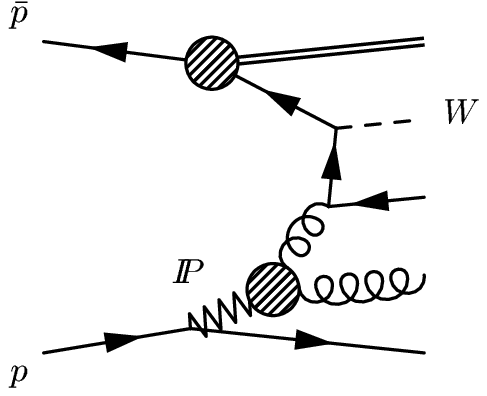}
\epsfig{file=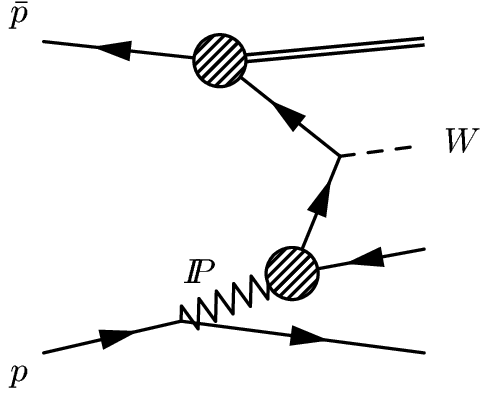}
\vspace*{5mm}
\caption{$W$ production in $p\bar p$ in the Pomeron model with a gluon (left) 
or a quark (right) from the Pomeron entering the hard subprocess.} \label{pomW}
\end{center}
\end{figure}

Diffractive $W$ production has been experimentally observed at the Tevatron by
the CDF collaboration at a relative rate $R_W^{\mathrm{CDF}}=(1.15\pm
0.55)\%$ \cite{CDF-W}. Only leptonic decays of the $W$'s are considered here,
since they
are easier to reconstruct due to a lower background. The interpretation of
diffractive $W$ production in the soft color exchange models is illustrated in
Fig.~\ref{fig:Wsci}. In order to have a leading proton, it is necessary to have
a gluon-initiated process, \ie, taking a gluon from a beam (anti)proton. The
color octet charge of the remnant can then be neutralized by a soft gluon
exchange between this remnant and some other color charge in the event. This
may be described in a parton basis as in the SCI model or in a string basis as
in the GAL model. In any case, this gives the possibility to produce a small
mass leading system, \eg, a single proton, separated by a rapidity gap to the
central system containing the $W$. 

In order to produce a leading proton, a parton with not too large
energy-momentum fraction $x$ from one beam proton will interact with a parton
from the other beam particle. Because of the small energy loss ($x$) from the
leading proton and the large mass of the $W$, the parton from the other beam
hadron will have to be quite energetic and is therefore typically a valence
quark. This also implies that the $W$ predominantly emerges in the hemisphere
opposite to that of the gap or the leading proton. These effects in our Monte
Carlo simulation produce the same correlations of rapidities and $W$ charge as
observed by CDF and used in the experimental analysis. 

In Pomeron models, on the other hand, $W$ production can be described, as
originally proposed and calculated in \cite{Bruni+I}, by the processes in
Fig.~\ref{pomW}. As discussed above, one folds a Pomeron flux from one of the
initial hadrons with a hard Pomeron-proton collision using parton densities in
the Pomeron. Since the charge-rapidity correlations are essentially of
kinematical origin, they also appear in this model. 

\begin{figure}[tp]
\begin{center}
\vspace*{-20mm}
\epsfig{file=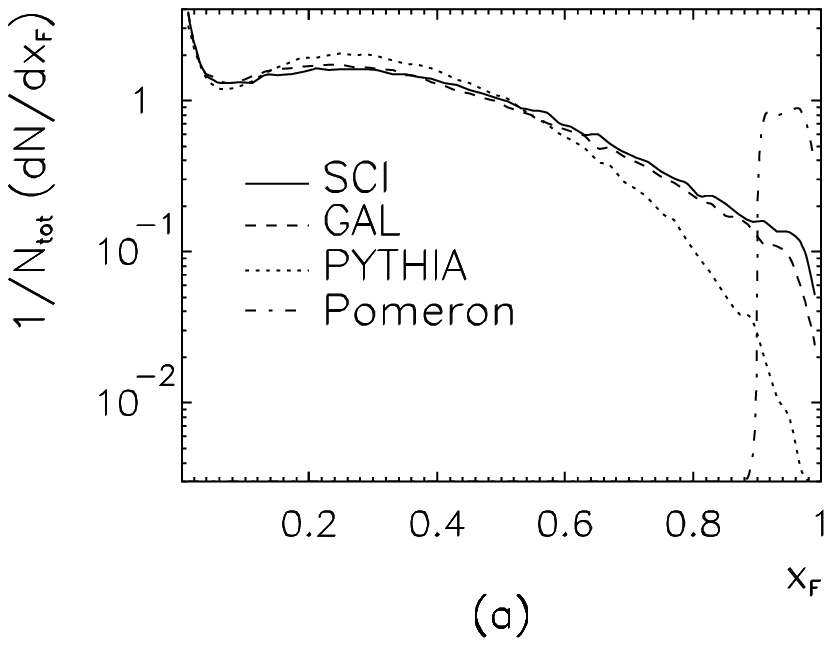} \\ 
\vspace*{-18mm}
\epsfig{file=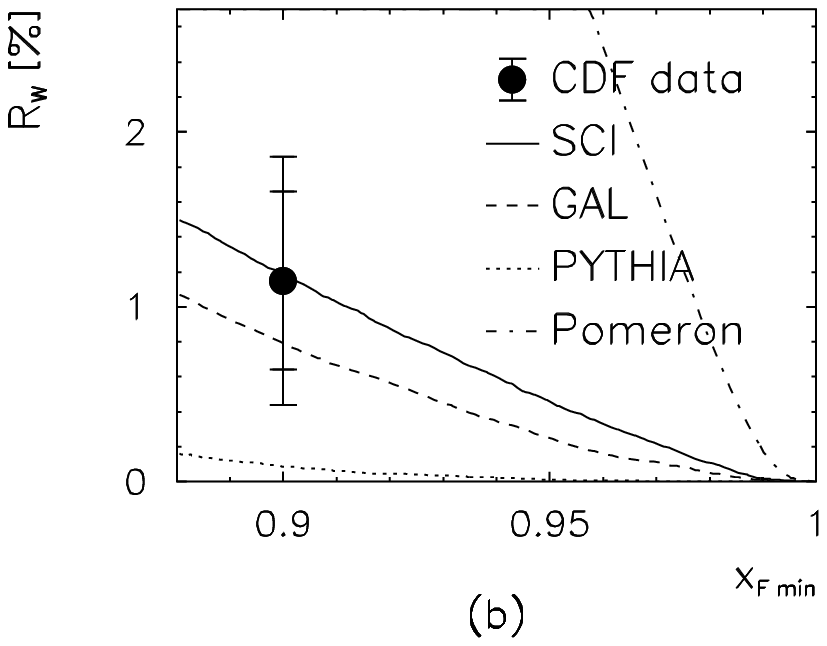} \\
\vspace*{-18mm}
\epsfig{file=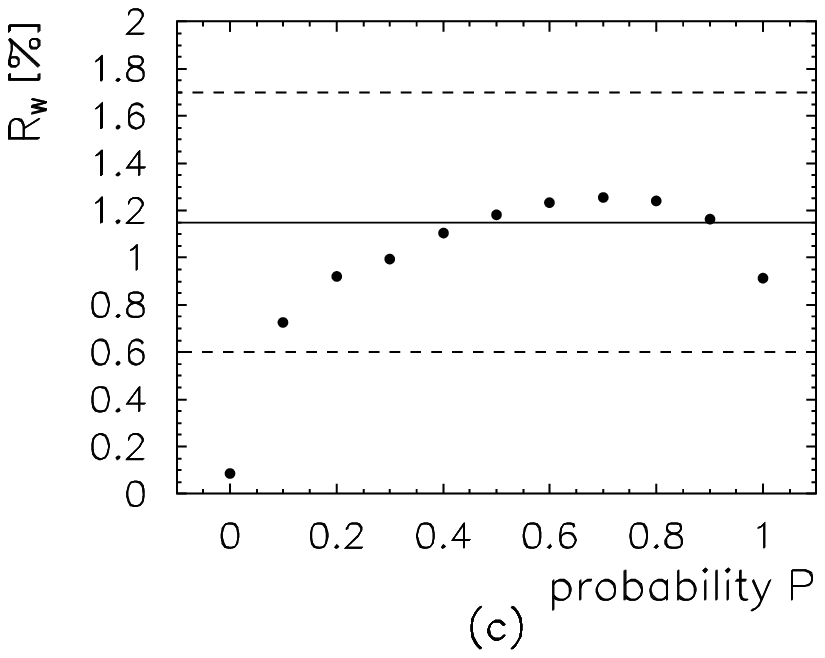}
\vspace*{-2mm}
\caption{Results from simulations of $W$ production in \ppbar \ collisions at
$\sqrt{s}=1.8$ TeV using different Monte Carlo models: default \Pythia ,
\Pythia \ with the soft color interaction models SCI and GAL added, and 
\Pompyt \ for Pomeron exchange. (a) Distribution in $x_F$ of leading 
(anti)protons. (b)
Relative rate $R_W$ of $W$ events with a leading proton having
$x_F>{x_F}_{\mathrm{min}}$. (c) Dependence of $R_W$ on the SCI probability
parameter $P$. The measured value of CDF \protect\cite{CDF-W} (with statistical
and systematic errors) corresponding to ${x_F}_{\mathrm{min}}=0.9$ is included 
as a point in (b) and as lines (central value and errors) in (c).}\label{plotW}
\end{center}
\end{figure}

The main results of our $W$ simulations are shown in Fig.~\ref{plotW}b which
shows that the SCI and GAL models reproduce the rate of diffractive $W$ as
observed by CDF, whereas the Pomeron model result is far above (about a factor
six) and standard \Pythia \ is much below the measured value. Here one should
remember that the SCI and GAL models are {\em not} adjusted to these data, but
have an absolute normalization which is fixed by the rate of rapidity gaps in
DIS at HERA, as discussed in Section \ref{sec-model}. This ability to reproduce
these CDF data is related to the increased rate of high-$x_F$ protons as shown
in Fig.~\ref{plotW}a. The Pomeron model, which is only applicable for 
$x_F\gtrsim 0.9$, overshoots the Tevatron diffractive $W$ rate 
if taken directly over from
its tuning to diffractive HERA data. As discussed in Section \ref{sec-pomeron},
this problem can be cured by introducing some essential modification of the
Pomeron model. Since the Pomeron model only applies in a limited $x_F$ range,
the curve in Fig.~\ref{plotW}a cannot be normalized to unit area and is instead
normalized based on its absolute cross in relation to the other models.
Concerning this Pomeron model curve, one should note that it is quite flat. The
basic $1/(1-x_F)$ dependence in the Pomeron model is here strongly distorted by
the kinematical suppression for $x_F\to 1$ imposed by the $W$ mass. This
implies that the cross section for diffractive $W$, as opposed to inclusive
single diffraction, is quite sensitive to the cutoff $x_{F\, min}$.

As pointed out, the rate of diffractive events may depend on whether they are
defined in terms of a gap or a leading proton. This CDF result is based on the
observation of a gap, but is essentially equivalent to requiring a leading
proton with $x_F>0.9$. The mild, essentially linear
variation of the SCI and GAL model results with $x_{F\, min}$ shown in
Fig.~\ref{plotW}b, demonstrate that the exact $x_{F\, min}$ value is not
crucial; in particular in view of the presently rather large error bars on the
experimental ratio $R_W$. 

It is now interesting to investigate how variations in the models affect the
results. To start with, we find that there is almost no discernible difference
between the results from the two variants of the SCI model, \ie, the one which
allows color reconnections between any pair of partons and the one which
requires one of the interacting partons to be in the remnant. This is because
practically all rapidity gaps come from reconnections involving a parton in the
remnant, representing the color background field. Color exchanges between the
more centrally produced partons from the hard scattering do not give rise to 
large gaps between the central and the leading systems. 
Consequently, in all simulations in this paper, the standard version of SCI is 
taken as the one involving at least one parton in the remnant.

One may ask how the diffractive ratio depends on the soft color exchange
probability $P$ for the SCI model or $P_0$ for the GAL model. Qualitatively, if
$P$ is large there will be more color reconnections, increasing the rate of gap
events. However, an increasing number of color exchanges may also destroy gaps,
through the possibility of reconnecting strings `across' an already formed gap.
This behavior is indeed found in the simulation and shown in Fig.~\ref{plotW}c
obtained using the SCI model. As can be seen, there is only a quite weak
dependence on $P$ as long as it does not approach its limiting values 0 or 1.
In accordance with earlier studies \cite{unified}, we take $P=0.5$ as our value
for the SCI probability. For the GAL model, we use the original value $P_0=0.1$
\cite{GAL} as discussed in Section \ref{sec-GAL}.

The improved model for sea quark treatment, which assigns some dynamics to the
sea quark partner in the case of scattering off a sea quark in the proton,
should be of relevance. The reason is that a sea quark may be viewed as coming
from $g\to q\bar{q}$ and thereby be like a gluon-induced process giving
diffractive $W$ production as discussed above. As described in Section
\ref{sec-remnant}, there are two variants of this sea quark treatment, SQT1 and
SQT2. Using one or the other, or neglecting this sea quark treatment, gives
somewhat different diffractive $W$ rates as shown in Table~\ref{tab-RW}, 
but the results are all within the experimental error. It could be argued that
SQT2 is more correct since it uses sea quark parton distributions to assign
momenta, while SQT1 uses pQCD parton splitting functions in the nonperturbative
region. Together with the fact that SQT2 gives slightly better agreement with
data, this is the preferred version that we use as standard. 

The multiple interaction model discussed in Section \ref{sec-remnant} has an
important influence on the results. It is clear that in Pomeron models,
additional parton-parton scatterings in an event would destroy any gaps, and
therefore the existence of a gap would signify that there were no such extra
scatterings in the event. In contrast, multiple interactions do not exclude
gaps in the soft color interaction models. In fact, the gap ratios shown here
for the SCI and GAL models include multiple interactions. The gap rate does
depend on the amount of multiple interactions, but switching them off only
leads to the somewhat increased gap rate shown in Table~\ref{tab-RW} which is
still consistent with the observed $R_W$. Therefore, at this stage of accuracy,
the multiple interactions do not present a problem. They must, of course, be
included at some level in order to reproduce various characteristics of the
underlying event. As discussed above, we have found a slightly increased value
of the basic transverse momentum cutoff parameter for these additional
parton-parton scatterings to avoid double counting of soft phenomena.

\begin{table}
\caption{Ratio $R_W$ of diffractive $W$ production obtained from different
variations of the models: sea quark treatment (SQT), multiple interactions
(MI), parametrization of parton densities in the proton (CTEQ) and in the
Pomeron (GS). Results from standard version 
models are shown in boldface.} \label{tab-RW}
\begin{tabular}{ll}
Model                & $R_W$ (\%)\\
\hline
{\bf SCI} incl.\ SQT2, MI                 & {\bf 1.2} \\
\quad '' \quad changing to SQT1           & 1.7       \\
\quad '' \quad switching off SQT          & 0.9       \\
\quad '' \quad switching off MI           & 1.7       \\
\quad '' \quad switching to CTEQ4L        & 1.0       \\
\hline
{\bf GAL}                                 & {\bf 0.8} \\
\quad '' \quad switching to CTEQ3L        & 1.0       \\\hline
{\bf Pompyt} GS I   & {\bf 7.2} \\
\quad '' GS II  & 11.6      \\
\hline
Default \Pythia      & 0.1       \\  
\end{tabular}
\end{table}

Finally, we have checked the dependence on the choice of parton distribution
parametrizations, and we find that the diffractive ratios are slightly smaller
(about 15--20\%) with CTEQ4L than with CTEQ3L. These variations of the SCI and
GAL models result in changes of the diffractive ratios (Table~\ref{tab-RW})
which illustrate the uncertainty of the models. We note that these variations
are all within the errors of the present experimental results.

In contrast to the soft color exchange models it is, as already emphasized, not 
possible to take the Pomeron model directly from HERA and use it to reproduce
the Tevatron data. The best result is achieved using model I for the parton
densities in the Pomeron, which results in a diffractive $W$ ratio which is six
times too large, whereas model II gives a rate about ten times too large (see
Table \ref{tab-RW}). Other parametrizations of the Pomeron parton densities
exist, but using them will not essentially change this disagreement with
Tevatron data which is also compatible with other investigations \cite{Alvero}.
We find, however, that some general characteristics such as the $\eta$
distributions of particles in an event, are the same for \Pythia{} with SCI and
for \Pompyt .

At this point, having examined variations of the models, we make an important
observation: the measurement of diffractive $W$ production was only made for
the leptonic decay channel $W \to e \nu$. When the $W$ instead decays to quarks,
these quarks must also be included in the soft color interactions since, given
the short $W$ lifetime, they are produced in a very small space-time region
embedded in the color background field of the colliding hadrons. This gives the
possibility that reconnections with these decay quarks rearrange  the color
structure of the event and destroy rapidity gaps. Therefore, the probability
for a diffractive event can be lower for hadronic than for leptonic $W$ decays.
This effect could be seen as an {\em apparent} change in the branching ratios
of $W$ decays, so that in a diffractive sample of events there will be a higher
branching ratio to leptons and a lower branching ratio to hadrons than what is
observed in the total, inclusive sample. In Pomeron models on the other hand,
no such effect should be present since the hard scattering is independent of
the gap-formation process. This has been confirmed by simulations with
\Pompyt{}.

The real branching ratios for $W$ are $B(W \to l\nu) = 32.2 \%$ and $B(W \to
q\bar{q}') = 67.8 \%$, and thus $B(W \to l\nu) / B(W \to q\bar{q}') = 0.475$.
Now, using the SCI model, but with both the leptonic and the hadronic decay
channels of the $W$ included, we find 
\begin{equation}
\left. \frac{B^{SCI}(W \to l\nu)}{B^{SCI}(W \to q\bar{q}')}
\right|_{\mathrm{diffractive}} =
\frac{39 \%}{61 \%} = 0.63 > 0.475.
\label{BR}
\end{equation}
Thus there are indeed different apparent branching ratios in the biased
diffractive $W$ sample. This is also reflected in the diffractive ratio $R_W$,
which drops from 1.2 to 1.0 when including hadronic $W$ decays.

Naively we would expect the same effect in the GAL model, but this is not
observed in our simulations. The reason is that for reconnections with the
decay
products of the $W$ the price in terms of increased string area is too large.
The quarks from the $W$ decay will form a separate color singlet system, which
is central in rapidity. Reconnecting this string with a string from a more
noncentral parton will typically mean an increase in area, which is strongly
suppressed in the model. Therefore we do not observe any shifted apparent
branching ratios in the GAL model, only in the SCI model.

The CDF paper \cite{CDF-W} also contains a study of the jet structure of
diffractive $W$ production. Only 8 out of 34 diffractive events were observed
to have a jet giving the ratio 24\%\ , but the relative error is large 
because of the low statistics. This fraction was used to estimate the quark
and gluon content of the Pomeron, and it was found that the measurement was
consistent with a quark dominated Pomeron (although the measured value of $R_W$
favors a gluonic Pomeron). An SCI model interpretation is also quite in order,
since we have verified that it can reproduce this measured rate of jets in
diffractive $W$ events. Here $W$ production with pQCD corrections in terms of
next-to-leading order tree level matrix elements and parton showers was
employed, however, the description turns out to be equally good using only LO
matrix elements and parton showers. 

Before moving on to other processes, we will briefly consider diffractive
$Z$ production, as this should be qualitatively similar to the $W$ case. This
has not been observed experimentally yet since the cross section and branching
ratio to leptons are both smaller for $Z$ than for $W$. We predict diffractive
ratios $R_Z$ that are smaller than the corresponding $R_W$ (see Table
\ref{tab-gapratios}):\ we get $(R_Z/R_W)_{\mathrm{SCI}}=0.83$ and
$(R_Z/R_W)_{\mathrm{GAL}}=0.64$. This difference is essentially accounted for by
the mass difference; it takes more energy to produce a $Z$, so there will be
less energy available for the leading proton, which will on average have a
lower $x_F$. Thus $R_Z$ will be lower than $R_W$. We have checked this by a
simulation where the $Z$ mass was set equal to the $W$ mass, resulting in a
ratio consistent with unity for SCI. We find similar results in the
Pomeron model, as expected based on general kinematical mass effects. 

In the GAL model, however, the suppression of $Z$ compared to $W$ is larger.
Simulating with the GAL model and the $Z$ mass changed to $m_W$, we get
$(R_Z^\prime/R_W)_{\mathrm{GAL}}=0.8$. Hence the larger mass is not the whole
reason. The difference between the SCI model and the GAL model is larger for
$Z$ than it is for $W$. This gives an indication that the dependence on the
hard
scale is different between the two models, as will be discussed in more detail
later.

To summarize this subsection, we have demonstrated that the SCI and GAL models
can indeed reproduce experimental data on diffractive $W$ production, while the
Pomeron model cannot without modifications. We have also studied some
variations of the models, and found that the `best model' is the same model as
the one used to reproduce diffractive HERA data, namely, SCI or GAL together
with the new model for sea quark treatment (SQT2), but here also with the
multiple interaction model necessary for {\ppbar{}} collisions.

 We have also pointed out some differences between the SCI, GAL and Pomeron
models, which could be used to experimentally discriminate between them. An
interesting such observable is the phenomenon of different apparent $W$
branching ratios in diffractive events.

\subsection{Diffractive beauty production}

\begin{figure}[t]
\begin{center}
\epsfig{file=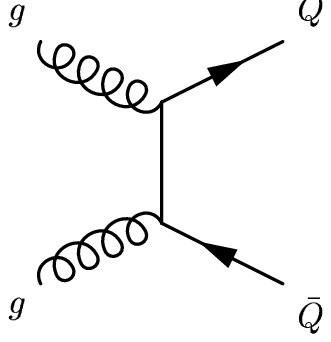, clip=, bbllx=22, bblly=690, bburx=109, bbury=796}
\quad
\epsfig{file=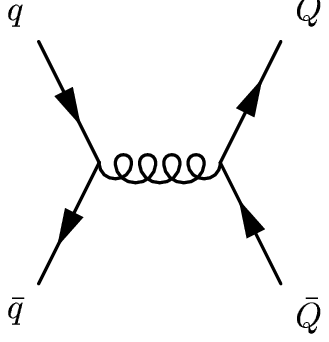, clip=, bbllx=22, bblly=690, bburx=109, bbury=796}
\quad
\epsfig{file=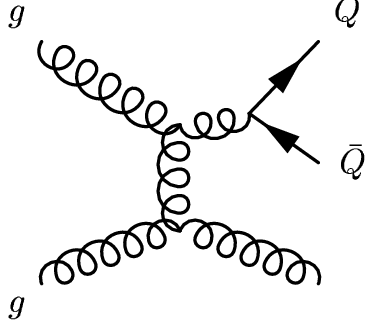, clip=, bbllx=22, bblly=690, bburx=119, bbury=796}
\vspace*{5mm}
\caption{Examples of pQCD processes for $Q\bar Q$ production: the left and 
middle diagrams show the two leading
order ($\alpha_s^2$) processes and the right diagram shows the most 
important next-to-leading order
($\alpha_s^3$) diagram.} \label{bbdiag}
\end{center}
\end{figure}

CDF has also measured diffractive \bbbar{} production in terms of open beauty 
in events with rapidity gaps, defined in the same way as in the $W$ case. The
resulting ratio of diffractive beauty production is $R_{b\bar b}=(0.62\pm
0.25)\%$ \cite{CDF-B}. 

In contrast to $W$ production, the description of heavy quark production needs
to include also higher order diagrams. In leading order (LO) pQCD heavy-quark
production occurs through $gg\to b\bar{b}$ and $q\bar{q}\to b\bar{b}$,
Fig.~\ref{bbdiag}ab. However, higher order processes involving gluon splitting
$g\to b\bar{b}$ are important. For example, the process $gg\to gb\bar{b}$
illustrated in Fig.~\ref{bbdiag}c gives a large contribution because it is an
$\alpha_s$ correction to the large cross section for gluon scattering ($gg\to
gg$). Matrix elements with explicit heavy-quark mass are available up to
next-to-leading order (NLO), but still higher orders may contribute at collider
energies. These can only be taken into account through the parton shower (PS)
approach which, although being approximate, has the advantage of resumming
leading logarithms to all orders.

We therefore investigate beauty production both in leading order and in higher
orders (HO) using \Pythia . The LO matrix elements include the $b$-quark mass
$m_b=4.5$ GeV. The higher orders are obtained through $g\to b\bar{b}$ in the
parton showers added to all LO $2\to 2$ QCD processes, except those producing 
$b\bar{b}$. The LO and HO contributions can then be added with their respective
cross section weights. The higher orders are tree level diagrams, whereas
virtual corrections are not taken into account in this approximation. 

\begin{figure}[t]  
\begin{center}
\epsfig{width= 0.49\columnwidth, file=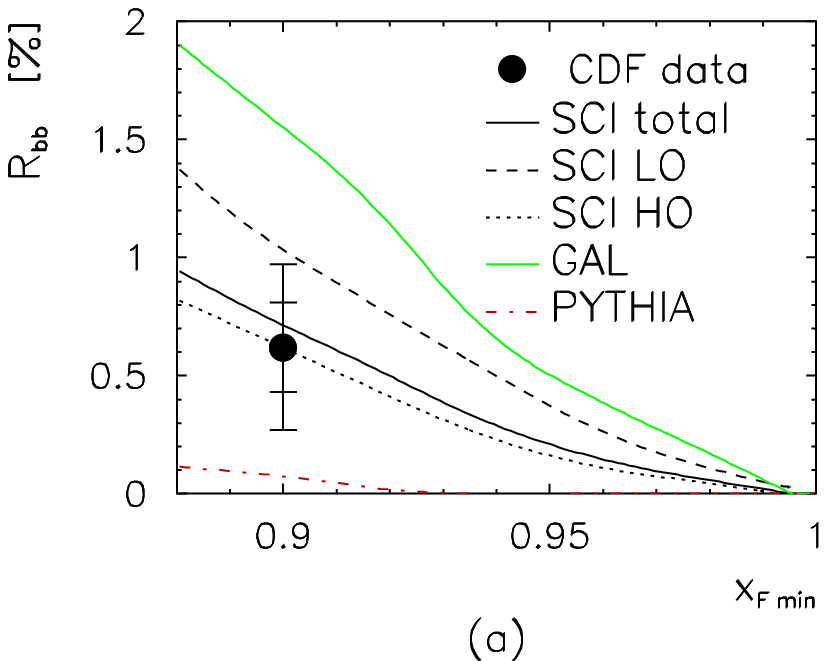} 
\epsfig{width= 0.49\columnwidth, file=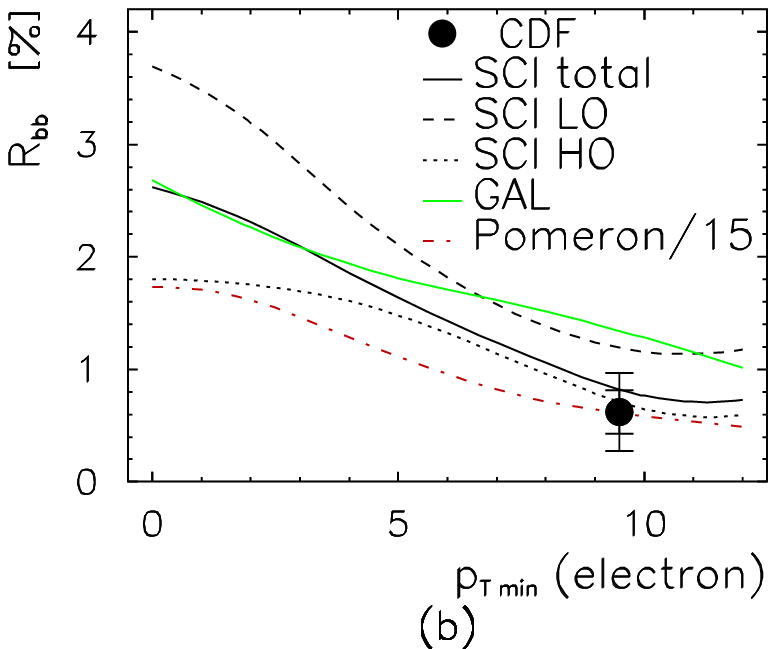} 
\caption{Relative rate of diffractive $b\bar b$ production in \ppbar \
collisions at $\sqrt{s}=1.8$ TeV as a function of 
(a) the minimum leading proton $x_F$, and 
(b) the minimum transverse momentum of the electron from the $b$-decay.
The CDF measurement \protect\cite{CDF-B} (with statistical and systematic 
errors) is compared to different models: default \Pythia , SCI with 
leading order and higher order contribution, GAL and in (b) also 
Pomeron exchange 
in \Pompyt \ scaled down by a factor 15.}
\label{bb_R}\label{bb_ptslope}
\end{center}
\end{figure}

The diffractive ratios obtained in this way are listed in Table
\ref{tab-gapratios} and are plotted as functions of ${x_F}_{\mathrm{min}}$ in
Fig.\ \ref{bb_R}. The separate LO and HO contributions in the figure show that
the LO gives a larger gap ratio, but the HO gives a larger contribution to the
total cross section. In contrast to $W$ production, GAL here gives a larger gap
ratio than SCI and is not in very good agreement with the experimental value.  
The SCI model gives excellent agreement as usual, whereas the Pomeron model is
a factor 15 too large compared to the measurement. 

Here we again note that the GAL model has a different energy dependence,
larger ratios for smaller hard scales and smaller ratios for larger hard
scales, as compared to the SCI model. 
This was already seen for $Z$ production
and we here anticipate the results from Sections \ref{sec-dij} and
\ref{sec-jpsi} and observe that the same holds for diffractive dijets and
$J/\psi$.

The experimental observation of $B$ mesons is based on electrons from their
decay. One requires these electrons to have a transverse momentum larger than
$p_{\perp \mathrm{min}}^e=9.5$ GeV. This is an important point, since we find
that the diffractive ratio $R_{b\bar b}$ depends on the value of $p_{\perp
\mathrm{min}}^e$, as shown in Fig.~\ref{bb_ptslope}. The three SCI curves shown
(LO, HO, and total) all have the same slopes. The Pomeron curve also has the
same slope, but as the absolute normalization is a factor 15 too large, it has
been correspondingly rescaled in the figure. The GAL curve is at the same level
as the SCI model for small $p_{\perp \mathrm{min}}^e$, but its slope is smaller
such that it overshoots the experimental data point. This different slope of
the GAL model is again a manifestation of its different scale dependence.

This dependence on $p_{\perp \mathrm{min}}^e$, which is effectively a
requirement on the transverse momentum of the $b(\bar b)$ quark, can arise from
an interplay of several effects. First, a higher $p_\perp$ requires larger
momentum fractions taken from the colliding protons, which means less energy
left for leading protons. Second, with higher $p_\perp$ the incoming
and outgoing partons radiate more, thus filling gaps. 
It is not a priori clear
how the underlying event affects this, but we have found that multiple
interactions do not change the slope of the curves, only the normalization.
Given these effects, one should realize that the measured diffractive beauty
ratio might be biased towards a lower value given the requirement of a
high-$p_\perp$ electron. 

\subsection{Diffractive dijet production}\label{sec-dij}

The process originally considered when introducing the concept of diffractive
hard scattering was jet production in high energy hadronic interactions
\cite{IS}. The transverse momentum ($p_\perp$) or transverse energy ($E_T$)
of the jets provides the hard scale necessary for the study of diffraction
based on a firm underlying parton picture. The experimental discovery by UA8 of
hard scattering phenomena in diffractive scattering was also in terms of events
with a leading proton and high-$p_\perp$ jets at the CERN \ppbar{} collider
\cite{UA8-1}. Additional UA8 data \cite{UA8-2} gave important results, which
were mainly interpreted in terms of the Pomeron model resulting in hard parton
density distributions in the Pomeron. 

Diffractive dijet production has also been observed by the CDF and D{\O}
experiments at the Tevatron. Initially, CDF observed \cite{CDF-JJ} events with
high transverse energy jets ($E_T > 20$ GeV) and a gap in the rapidity region
opposite to the dijets in $p\bar p$ collisions at $\sqrt s = 1800$ GeV, while
D{\O} has reported \cite{D0-JJ} observation of events with a similar topology
($E_T > 12$ GeV and $E_T > 15$ GeV) at the two center of mass energies 
$\sqrt{s}=630$ and 1800 GeV, respectively. The analyses are quite analogous 
to that for diffractive $W$ discussed above, with the observed gap equivalent
to a leading  proton with $x_F>0.9$. 

\begin{figure}[t]
\begin{center}
\epsfig{width= 0.3\columnwidth,file=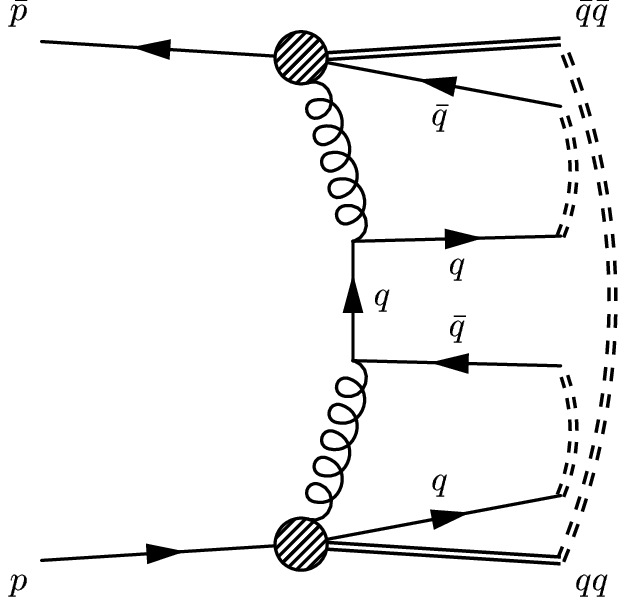}
\epsfig{width= 0.3\columnwidth,file=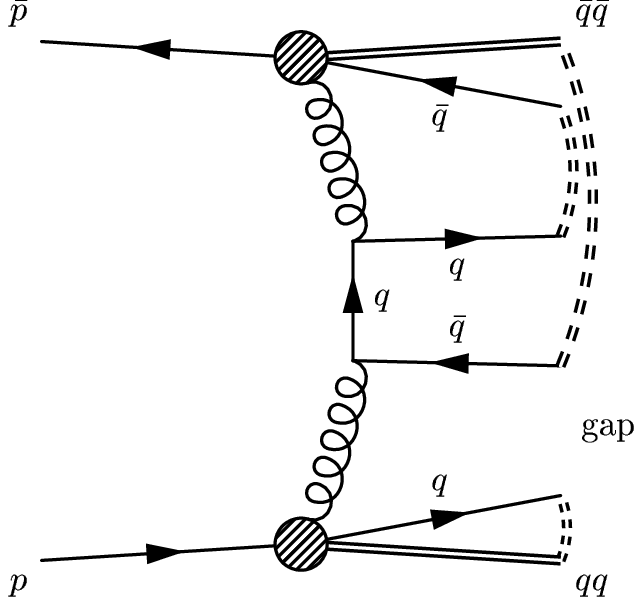}
\epsfig{width= 0.3\columnwidth,file=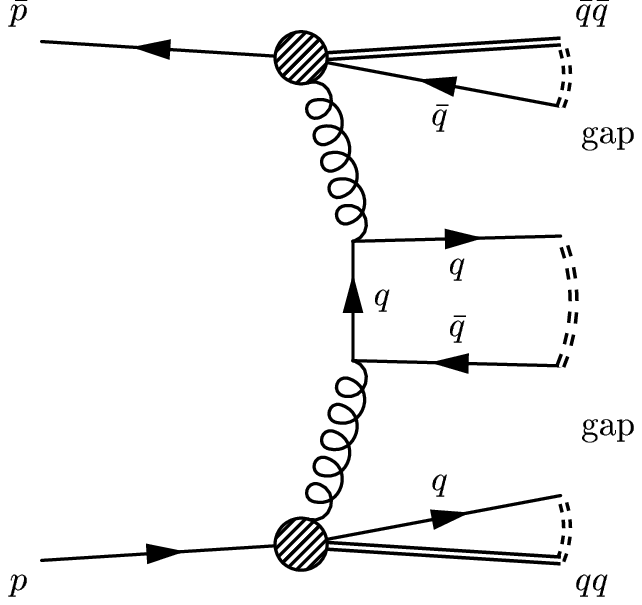}
\vspace*{5mm}
\caption{Dijet production in $p\bar{p}$ collisions with string topologies
(double-dashed lines) before and after soft color interactions resulting in
events with one gap (leading particle) and two gaps (leading particles).}
\label{pp-dijets}
\end{center}
\end{figure}

\begin{figure}[t]
\begin{center}
\epsfig{file=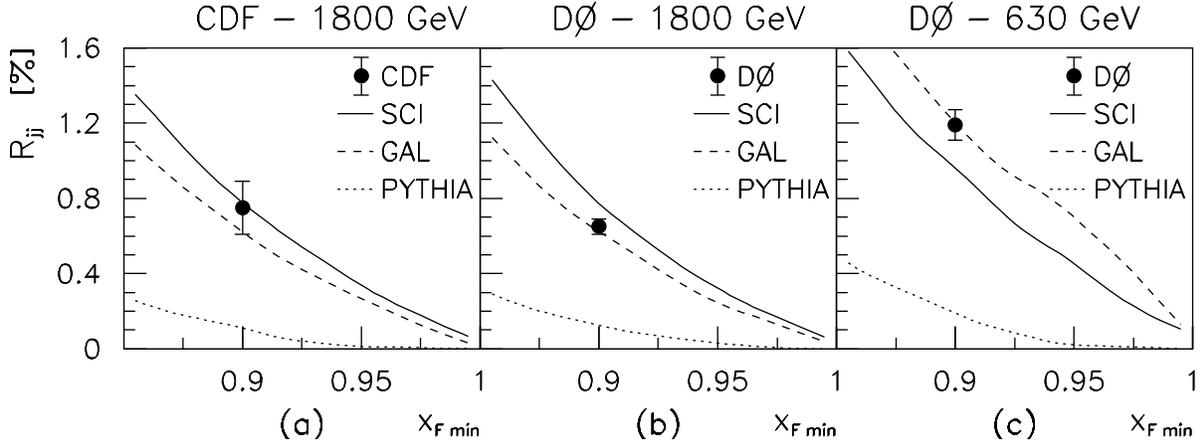}
\vspace*{5mm}
\caption{Relative rate of diffractive dijet production in $p\bar p$ collisions
as a function of $x_{F\, min}$, the minimum momentum fraction of the leading
proton; (a) CDF \protect\cite{CDF-JJ} and (b,c) D{\O} \protect\cite{D0-JJ} 
results at
different cms energies are compared to model
results (default \Pythia, SCI and GAL).}
\label{figjj}
\end{center}
\end{figure}

This kind of events occurs naturally in the soft color exchange models as
illustrated in Fig.~\ref{pp-dijets}. Applying the SCI and GAL models to jet
production in \Pythia , described by leading order QCD $2\to 2$ scattering
processes with parton showers added for higher orders, results in a good
description of the observed diffractive dijet ratios $R_{jj}$, as shown in
Table~\ref{tab-gapratios} and Fig.~\ref{figjj}. We emphasize that it is exactly
the same SCI and GAL models as used for diffractive $W$ and $b\bar{b}$ above,
only the hard subprocess has been changed. We have investigated the dependence 
of the results on the reconnection probability $P$, $p_\perp^{\mathrm{min}}$ 
in the multiple interaction model, different aspects of the sea quark 
treatment, and arrived at the same conclusions as for the $W$ case in 
Section~\ref{Wprod}.

The other models cannot reproduce the measured $R_{jj}$; default \Pythia{} is
far below data (Fig.~\ref{figjj}) and the Pomeron model is above (not shown
explicitly).  

CDF has recently presented a new sample of diffractive dijet events, where the
signature of diffraction is a leading antiproton observed in Roman pot
detectors \cite{CDF-AP}. The reported results are based on events with
antiprotons in the range $0.905< x_F < 0.965$ and two jets with $E_T > 7$ GeV.
Since this offers a new testing ground for the models, we have investigated the
production of dijet events with a leading antiproton and compared the results
of the models with the observed CDF data. We note that CDF uses the variable
$\xi$ to denote the antiproton fractional momentum loss, which is related by
$x_F=1-\xi$ to the variable $x_F$ consistently used in this paper. 

\begin{figure}[t]
\begin{center}
\epsfig{width= 1\columnwidth, file=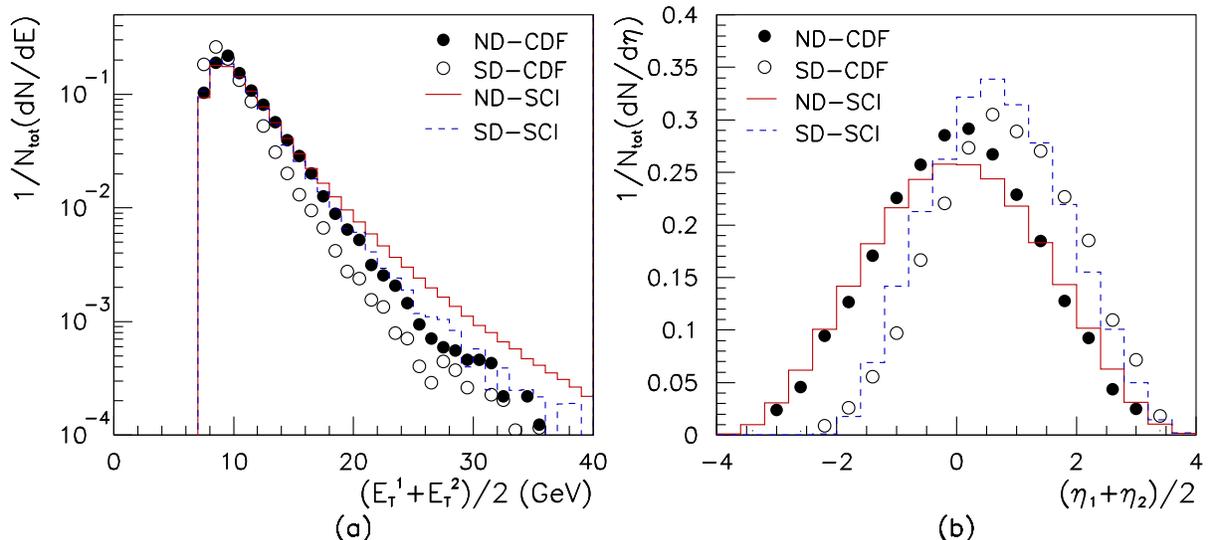} 
\caption{Distributions of (a) mean transverse energy and (b) pseudorapidity of 
the dijet system in nondiffractive (ND) 
and single diffractive (SD) $p\bar{p}$
events at $\sqrt{s}=1.8$ TeV. The points are CDF \protect\cite{CDF-AP} 
data and
the histograms are \Pythia{} with the soft color 
interaction (SCI) model added.}
\label{eta}
\end{center}
\end{figure}

In Fig.~\ref{eta} we compare characteristic features of the dijet systems in our
models and in data.  The data show that the $E_T$ distribution of the
diffractive sample falls steeper than that of the nondiffractive sample. This
behavior is present in both the SCI and GAL models, although the exact shape is
not very well reproduced. This may be related to a mismatch between our jet
reconstruction procedure and the experimental one, or \Pythia{} being limited
to leading order matrix elements without next-to-leading order corrections for
the basic jet cross section. The rapidity distribution of the jets is in the
diffractive sample shifted into the hemisphere opposite to the leading
antiproton, a characteristic which is well described by both models, see
Fig.~\ref{eta}b for the case of SCI.

\begin{figure}[t]
\begin{center}
\epsfig{width= 1\columnwidth, file=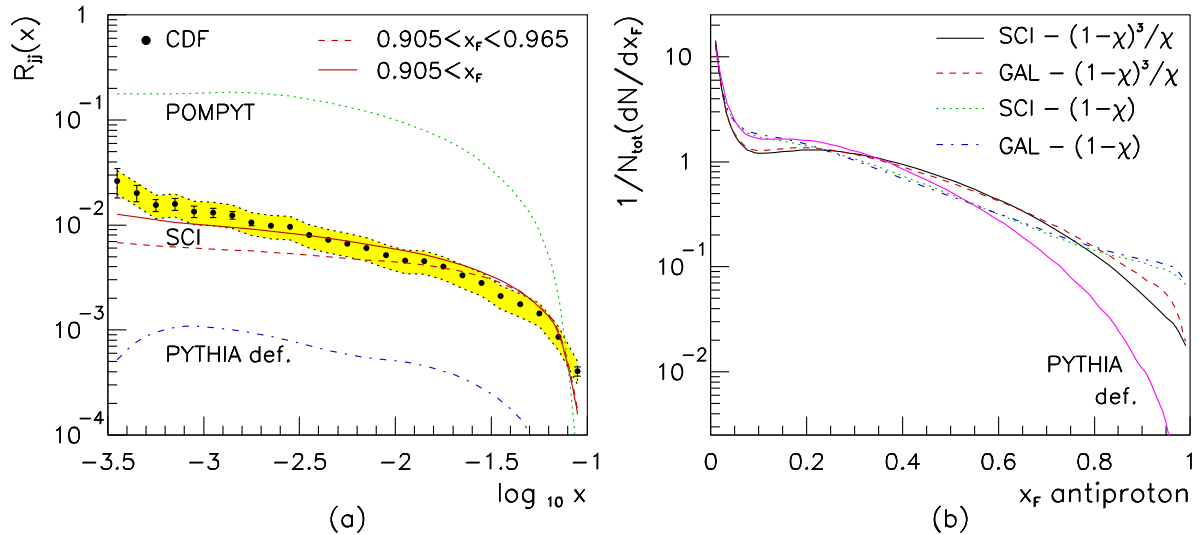} 
\caption{(a) Ratio of diffractive to nondiffractive dijet events versus 
momentum fraction $x$ of the interacting parton in $\bar{p}$. The points are 
CDF data \protect\cite{CDF-AP} and the shaded band shows 
the $\pm 25 \%$ systematic normalization uncertainty. The curves are 
from the \Pompyt{} Pomeron model, default \Pythia{} and the SCI model 
(for two $x_F$ regions). 
(b) Distribution in $x_F$
of the leading $\bar{p}$ from default \Pythia{} and the soft color exchange
models (SCI, GAL) with varied modeling of the remnant.} 
\label{xf}
\end{center}
\end{figure}

CDF has furthermore extracted the ratio of diffractive to nondiffractive dijet
events as a function of the momentum fraction $x$ of struck parton in the
antiproton. This $x$ can be evaluated from the transverse energy and rapidity
of the jets using the relation
\begin{equation} 
x=\frac{1}{\sqrt{s}} \sum_{j=1}^{\mathrm{2\; or\; 3}} E_T^j ~ e^{-\eta^j}
\label{x-antip}
\end{equation}
where the sum includes the two leading jets, plus a third jet if it has 
$E_T > 5$ GeV. In Fig.~\ref{xf}a we compare their data with the results from the
models. The Pomeron model overshoots the data by an order of magnitude, while
default \Pythia{} is too low by a similar factor. The soft color exchange
models give a fairly correct description, reproducing the overall behavior and
giving the correct total ratio. Going into finer details, we note that as $x_F$
approaches unity ($x_F>0.965$), the slope of this ratio with $x$ becomes more
steep in the models (as seen in Fig.~\ref{xf}a, where this contribution is
included in the full curve). This behavior seems not to be quite in accord with
CDF results, which
indicate a constant slope as $x_F$ varies \cite{CDF-AP}. This dependence in the
model is mainly due to the details of the remnant treatment, which affect the
steepness of the ratio.

The measurement of the leading antiproton provides a test of exactly how the
beam particle remnant is handled in the model. In order to explore this we have
investigated the effects of the alternative remnant handling procedures
available in \Pythia{}. Since diffractive events arise dominantly in the SCI
and GAL models from gluon-induced processes, the remnant typically contains the
three valence quarks. As described in Section~\ref{sec-remnant}, this remnant
is split into a quark and a diquark taking energy-momentum fractions $\chi$ and
$1-\chi$, respectively. The probability distribution ${\cal P}(\chi)$ cannot be
deduced from first principles, but is given by some parametrization. As our
standard choice we use ${\cal P}(\chi) \sim (1-\chi)$, giving in the mean one
third of the remnant energy-momentum to the quark and two thirds to the
diquark. We have also tried other parametrizations, in particular the parton
distribution-like form ${\cal P}(\chi) \sim \chi ^{-1} (1-\chi)^3$. The
antiproton $x_F$ spectra obtained are shown in Fig.~\ref{xf}b. The SCI and GAL
results are quite similar, but both depend significantly on this remnant
treatment. Of course, the main effect in Fig.~\ref{xf}b is the large increase
of antiprotons at large $x_F$ when going from default \Pythia{} to the SCI or
GAL model resulting in an overall description of the diffractive rates. The
finer details of the diffractive events will, however, depend on the details in
the modeling of the remnant. 

Summarizing the investigation of diffractive dijets, the soft exchange models
do a very good job in
reproducing the overall ratios of diffractive to nondiffractive dijet
production. They also give a good agreement with the kinematical
distributions observed for this type of events. However, some detailed results
depend on the treatment of the proton remnant in the Monte Carlo. The new
diffractive Tevatron data based on a leading antiproton provide additional
tests of the models. 

\section{DPE -- `Double leading Proton Events'}\label{sec-dpe}

Related to single diffraction are events with {\em two} leading protons with
associated gaps. These protons are at the opposite extremes in phase space,
\ie, at $x_F\to +1$ and $x_F\to -1$, and their associated gaps are in the
forward and backward rapidity regions, respectively. In the Regge framework
these events are described by a process where the two beam protons each emit a 
Pomeron. These Pomerons then interact, producing a central system which is 
separated in rapidity from the two quasi-elastically scattered beam protons. 
This class of events has therefore been called double Pomeron exchange (DPE). 
This nomenclature is, however, based on an interpretation in a certain model 
and it would be better to classify them independently of any model and only 
based on their experimental signature. In order to keep the well established 
abbreviation DPE, we propose to call them `Double leading Proton Events'. 

These DPE events occur naturally in the soft color interaction models, where the
final color string topology may also produce two rapidity gaps as illustrated
in Fig.~\ref{pp-dijets}c. With one single mechanism for soft color exchanges,
different final states will emerge and can be classified in the same way as
experimentally observed events: no-gap events, single diffractive events with
one gap or a leading proton, or DPE events with two gaps or two leading
protons. It is therefore straightforward to extract such events from the Monte
Carlo simulations based on the SCI and GAL models. 

Both CDF~\cite{CDF-DP} and D{\O}~\cite{D0-DP} have observed such DPE events
having a dijet system in the central region. They were first identified by two
rapidity gaps, one in the forward and one in the backward region. The ratio of
two-gap events to one-gap events observed by CDF is well reproduced by the SCI
model, as can be seen in Table~\ref{tab-gapratios}. Although D{\O} has not made
such a ratio available, the expectation from the models would be of the same
magnitude ($\sim 0.2 \%$). Recently, CDF has reported DPE dijet events defined
by a leading antiproton and a rapidity gap on the opposite proton side
\cite{CDF-DPE}. In the data set of single diffractive dijet events with leading
antiproton, they have observed a subset with a rapidity gap on the outgoing
proton side at a rate given in Table~\ref{tab-dpe}. By studying the kinematical
correlations between a leading particle and the associated gap, CDF describes
the DPE events in terms of a leading proton with $0.97<x_F<0.99$, although no
such proton is actually observed. 

\begin{table}
\caption{Rates of DPE dijet events in data compared to SCI and GAL models;
relative to single diffractive dijet events and absolute cross section.}
\label{tab-dpe} 
\begin{tabular}{lcc}
 & $\tilde{R}^{DPE}_{SD}\;\;\; [\% ]$~\tablenotemark[1] 
 & $\sigma^{DPE}\;\;\; [nb]$ \\
\hline
CDF \cite{CDF-DPE} & $0.80 \pm 0.26$ & $43.6\pm 4.4 \pm 21.6 $ \\
SCI  & $0.54\; \pm 0.05$ & $5$~\tablenotemark[2] -- $25$~\tablenotemark[3] \\ 
GAL  & $0.44\; \pm 0.05$ & $6$~\tablenotemark[2] -- $40$~\tablenotemark[3] \\
\end{tabular}
\tablenotemark[1]{~Calculated per unit $x_F = 1-\xi$ of leading proton.}\\
\tablenotemark[2]{~Leading proton in $0.97<x_F<0.99$.}\\
\tablenotemark[3]{~Only gap requirement on proton side.}\\
\end{table}

Table~\ref{tab-dpe} also contains the results of the SCI and GAL models.
Applying the leading proton condition strictly results in too low cross
sections, but when instead using the more generous gap definition the models
reproduce the measured cross section within the errors. This difference between
the two approaches illustrates our warning above that leading particle and gap
definitions need not be exactly equivalent. In particular, experimental
smearing effects may become important when approaching the phase space limit
$x_F\to 1$.
The absolute cross section is more sensitive to details in the model, such as
the remnant treatment and the previously mentioned lack of NLO corrections in
\Pythia{} may also play a role. With the uncertainties in both data and models
in mind, one may conclude that the models give essentially the correct cross
section for DPE events. 

\begin{figure}[ht] 
\begin{center}
\epsfig{width= 0.45\columnwidth, file=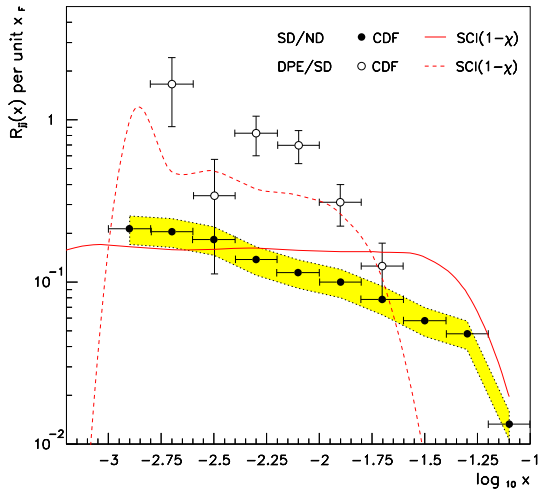} 
\caption{For dijet event samples, the ratio of DPE to single diffraction 
(per unit
$x_F^p$) and the ratio of single diffraction to nondiffraction (per unit
$x_F^{\bar p}$), as a function of the momentum fraction $x$ of the 
struck parton in $p$
and $\bar{p}$, respectively. The CDF data \protect\cite{CDF-DPE}, with 
statistical
errors and a $\pm 20\%$
normalization uncertainty band for SD/ND, is compared to the SCI model 
using the
$(1-\chi)$ parametrization for the remnant treatment in \Pythia .} 
\label{plot_r_dpe}
\end{center}
\end{figure}

This discussion illustrates the difficulty to exactly reproduce data in a Monte
Carlo model which is ambitious enough to attempt to describe the detailed
dynamics of nonperturbative QCD processes. This problem is accentuated for DPE
events, where the gaps and leading particles in both the forward and backward
region mean a stronger dependence on the details of the remnant treatment.
Using the remnant splitting ${\cal P}(\chi) \sim (1-\chi)$ based on simple
counting rules, the ratio of DPE to SD events and of SD to ND events gets
closer to the measured values than other options for ${\cal P}(\chi)$ provided
in \Pythia{}. The $x$-dependence of these ratios are shown in
Fig.~\ref{plot_r_dpe}. The curve for DPE/SD is obtained with the same leading
proton requirement as CDF derived from the observed rapidity gap. The SD/ND
ratio differs from the one in Fig.~\ref{xf}a by being calculated per unit
$x_F^{\bar p}$, which not only changes the normalization but also the slope.
The main features of the data are described by the SCI model, but there are
discrepancies related to the mentioned problems of the remnant treatment. The
main result in Fig.~\ref{plot_r_dpe} is, however, the breakdown of diffractive
factorization, which is quantified by the ratio of SD/ND to DPE/SD (=$0.19\pm
0.07$) being so clearly different from unity \cite{CDF-DPE}. This important
result also emerges from the models. 

\begin{figure}[t] 
\begin{center}
\epsfig{width= 0.33\columnwidth, file=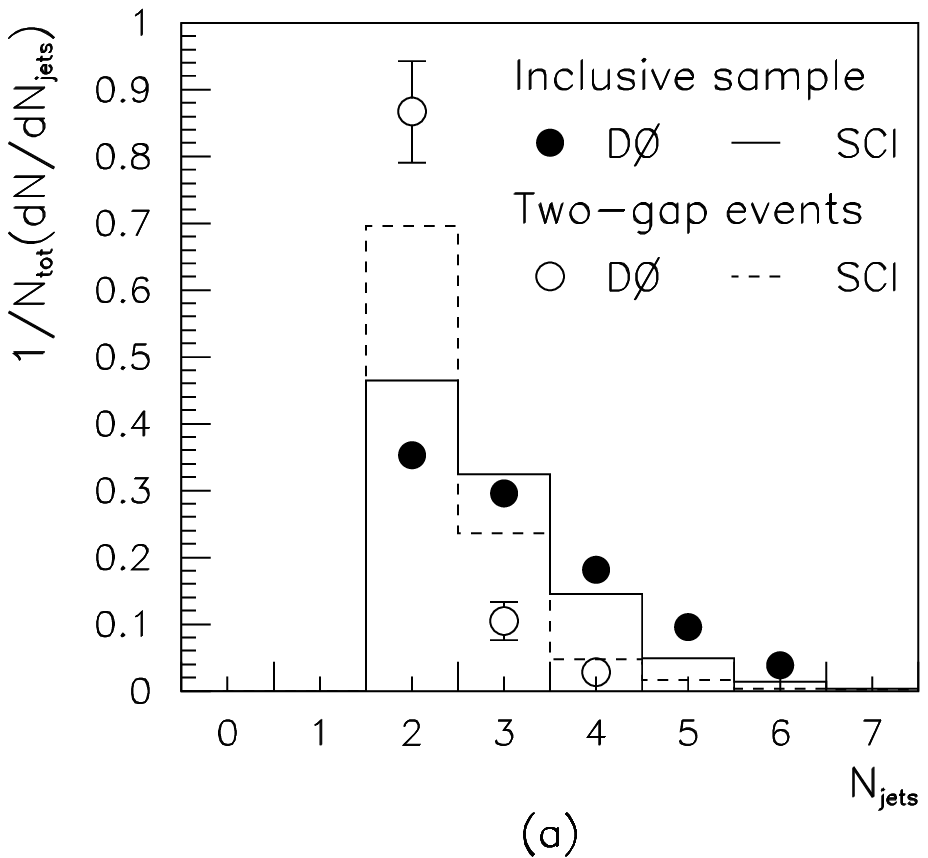} 
\epsfig{width= 0.66\columnwidth, file=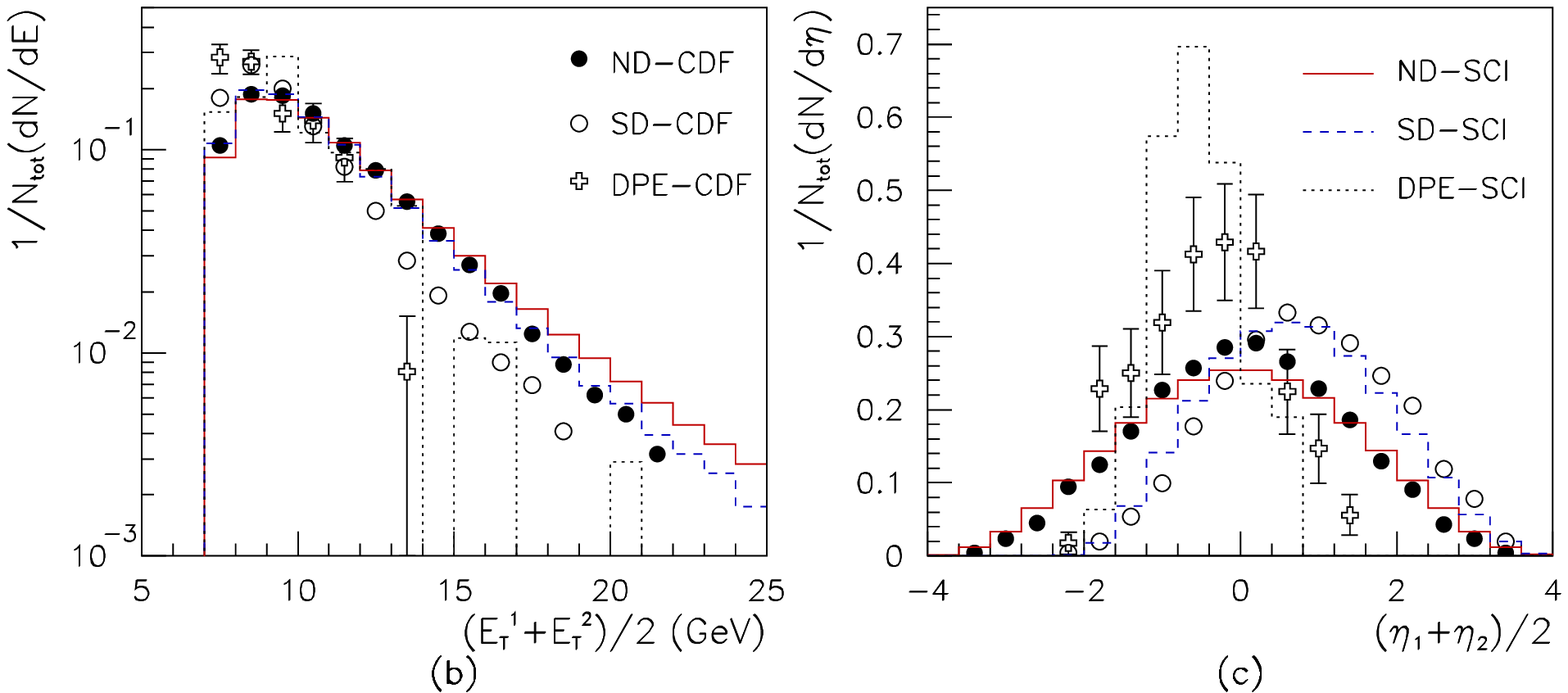} 
\caption{Comparison of jet properties in inclusive, nondiffractive (ND), single
diffractive (SD) and DPE events in \ppbar \ at $\sqrt{s}=1.8$ TeV; 
the points are D\O~(preliminary)\protect\cite{D0-DP} and 
CDF \protect\cite{CDF-DPE} data 
and the histograms are from the soft color interaction model. 
(a) Multiplicity of jets with $E_T>15$ GeV. 
(b,c) Distribution of mean transverse energy and pseudorapidity of the 
two jets with highest $E_T$ (dijet system).}  \label{dpe_jets}
\end{center}
\end{figure}

After this discussion of the rates of DPE events, we turn to some of their
internal properties. Fig.~\ref{dpe_jets} shows some essentials of the jets in
DPE events compared to inclusive and single diffractive events. Higher jet
multiplicities are clearly suppressed in DPE events compared to the inclusive
sample. The slopes of the jet-$E_T$ distributions have a tendency to increase
from nondiffractive to single diffractive to DPE events. This can be understood
by the limitations on the energy in the hard scattering subsystem due to
leading particle effects. The rapidity distribution, which is symmetric around
zero for nondiffractive events, is shifted when gap or leading proton
conditions are applied on either side. All these features are qualitatively
reproduced by the SCI and GAL models. Some discrepancies can, however, be found
in the details. The $E_T$ distributions in the models seem to have somewhat
too small slopes and higher jet multiplicities are not sufficiently suppressed 
in DPE events. These deficiencies may be due to a mismatch between data and 
model regarding the jet reconstruction or the lack of NLO corrections in the 
hard scattering matrix elements used in \Pythia .

We have shown in this section how soft color exchange models go beyond their
original purpose and explain more than just single diffraction; thus giving a
natural description of diffractive events with two gaps or corresponding
leading particles. The two leading particles imply an increased sensitivity to
the remnant treatment, providing possibilities to test and improve the details
of the Monte Carlo model.  

\section{Diffractive $J/\psi$ production}\label{sec-jpsi}

In the last section it was shown that the soft color interactions can produce 
two
rapidity gaps in the same event and thereby provide a description of DPE. In
this section we will demonstrate an even more striking effect where the soft
color interactions give rise to two different phenomena in the same event,
namely both a rapidity gap and turning a color octet \ccbar \ pair into a
singlet giving a $J/\psi$. The results of our models are predictions to be
tested against the data that should appear soon given the very recent
observation by CDF of such diffractive $J/\psi$ events. It will be a highly
nontrivial result if both the gap formation and the $J/\psi$ production can be
well explained with one and the same model for non-pQCD dynamics. 

To start with, let us leave diffraction aside and concentrate on the $J/\psi$
production. The main point here is that the soft color interaction, \eg, seen
as a soft color-anticolor gluon exchange, can change the color charge of a
\ccbar \ pair. A sizable fraction of the large cross section for pQCD
production of color octet \ccbar \ pairs can then be turned into color singlet
\ccbar . These will form onium states when their invariant mass is below the
threshold for open charm production. It is a remarkable fact \cite{SCI-onium}
that exactly the same SCI model that was used above, reproduces the observed
cross sections of high-$p_\perp$ charmonium and bottomonium in \ppbar \ at the
Tevatron. Since these cross sections are factors of ten larger than the
prediction of conventional pQCD in terms of the color singlet model, where the
\ccbar \ is produced in a singlet state, they need a radically new explanation.

\begin{figure}[t]
\center{\epsfig{width=1 \columnwidth,file=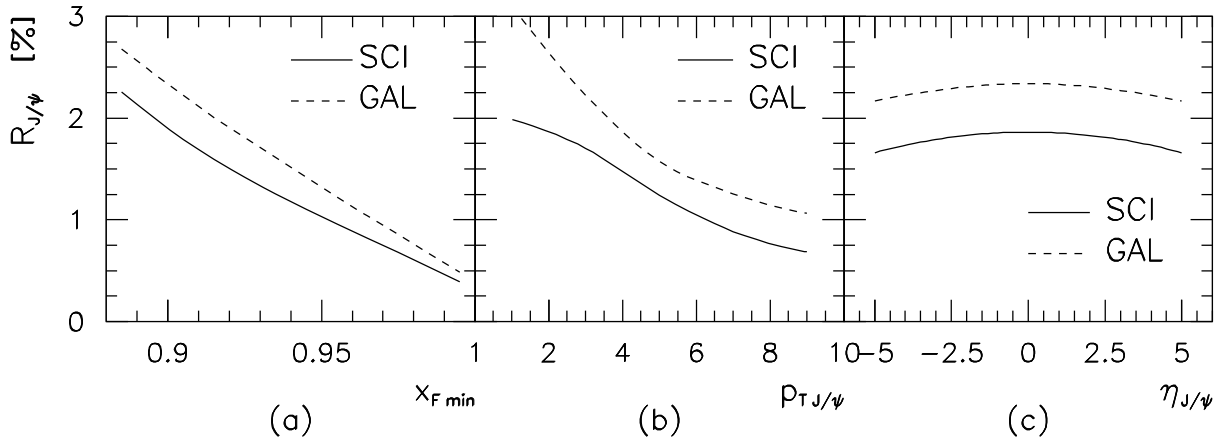}}
\caption{Predictions of the SCI and GAL models for the ratio of diffractive to
inclusive $J/\psi$ production in \ppbar \ at $\sqrt{s}=1.8$ TeV as a function
of (a) $x_{F\, min}$ of the leading proton, (b) the transverse momentum, 
and (c) the rapidity of the $J/\psi$ ($x_{F\, min}=0.9$ in (b,c)).}
\label{diffJpsi}
\end{figure}

The production of charmonium states in fixed target hadronic interactions at
different energies can also be described by these kinds of soft color
interaction models, as demonstrated in \cite{Cristiano}. Furthermore, elastic
and inelastic photoproduction of $J/\psi$ at HERA has been investigated from
the perspective of soft color exchanges \cite{johan-quarkonium}. Although SCI
and GAL show good agreement with data for the energy dependence of the cross
section, the normalization is uncertain since these models are based on leading
order matrix elements. The results are, therefore, sensitive to the choice of
factorization and renormalization scale, and in the elastic case, the treatment
of the proton remnant. 

Given this success of soft color interaction models to describe inclusive heavy
quarkonium production, we now turn to diffractive $J/\psi$ production at the
Tevatron. The predictions of the SCI and GAL models are shown in
Fig.~\ref{diffJpsi}. The ratio of diffractive to nondiffractive $J/\psi$ events
is in the range 1--2\%, depending on $p_{\bot}$ and $\eta$ of the $J/\psi$.
This predicted  ratio seems to be in agreement with the recent preliminary CDF
result \cite{Goulianos} of $(0.64 \pm 0.12)/{\cal A}$, where ${\cal A} \sim
0.4$ is an estimated rapidity gap acceptance. These diffractive events are
experimentally defined as events with a rapidity gap and we have performed the
analysis similarly to the aforementioned hard processes with a rapidity gap.

For production of $c\bar c$ with appropriate invariant mass to form $J/\psi$,
we found that higher order contributions are very important, which was also
demonstrated in \cite{SCI-onium}. The leading order
production through $gg \rightarrow c\bar{c}$ and  $q\bar{q} \rightarrow
c\bar{c}$ (Fig.~\ref{bbdiag} a,b) are included through massive matrix elements,
while the higher order tree level contributions are taken into account 
approximately
through the parton shower approach (main contribution in Fig.~\ref{bbdiag}c).
LO and HO give the same ratio of diffractive to nondiffractive $J/\psi$ when
considered independently, but the HO mechanism gives a higher absolute cross
section and therefore dominates the diffractive $J/\psi$ events. 
  
To conclude, we find that the ratio of diffractive to nondiffractive $J/\psi$
predicted in the SCI and GAL models seems to be in agreement with expectations
based on recent preliminary experimental results. This shows that the same soft
color interaction mechanism can be used to describe both gap formation and
quarkonium production, even occurring in the same event!

\section{Conclusions}\label{sec-conclusions}

A proper understanding of nonperturbative QCD has not yet been possible
based on rigorous theory. The development of phenomenological models is
therefore a useful approach. By considering soft effects in hard scattering
events one can have a firm basis in terms of a parton level process which can
be calculated in perturbation theory. Below the cutoff for the perturbative
treatment, further interactions occur abundantly because of the large coupling
$\alpha_s$ at small scales. The problem is then to model these soft
interactions properly. The soft interactions can have large effects on the
hadronic final state. This was demonstrated in Fig.~\ref{plotmaxgapsize}, where
frequently occurring large rapidity gaps on the parton level were filled
through the hadronization process resulting in a strong, exponential
suppression of large gaps at the hadron level. Conventional hadronization
models, like the Lund string model, have a substantial theoretical input and
describe very well many aspects of the hadronic final states. Nevertheless,
they are still not derived from fundamental QCD theory, but are of
phenomenological character and depend on
which data have been considered when constructing them. The models may
therefore need the introduction of new aspects or new dynamics as other data or
new observations are considered. 
 
The soft color interaction approach investigated in this paper is an example of
such new dynamics. We have argued that these interactions are a natural part of
the process in which bare perturbative partons are dressed into
nonperturbative ones and of the formation of color flux tubes between them. In
the SCI model this may be viewed as the perturbatively produced partons
interacting softly with the color medium of the proton as they propagate
through it. Interactions of a color charge with a color background field is a
more general problem which has been investigated using other theoretical
approaches and received increasing interest in recent years. Examples of
effects considered are large $K$-factors in Drell-Yan processes and synchrotron
radiation of soft photons \cite{Nachtmann} as well as diffractive DIS in a
semiclassical model \cite{Buchmueller}. The new approach to diffraction in
\cite{Dino-new} may also be possible to interpret in a soft color interaction
scenario. 

Our phenomenological approach is formulated in terms of the SCI and GAL models
which are added to the well-known Monte Carlo programs \Lepto \ and \Pythia . A
new stage of soft color interactions is introduced after the conventional
perturbative processes, described by matrix elements and parton showers, but
before applying the standard Lund string hadronization model. The SCI model is
formulated in a parton basis, with soft color exchange between quarks and
gluons, whereas the GAL model is formulated in a string basis, with soft color
exchange between strings. In both cases, this causes a change of the color
string topology of the event such that another hadronic final state will result
after hadronization. These fluctuations will sometimes result in a region where
no string is stretched giving a rapidity gap after hadronization. In both
models there is only one new parameter, giving the probability for such color
exchanges. The value of this parameter is chosen such that the rate of
diffractive rapidity gap events observed in DIS at HERA is reproduced. 

The main result of this paper is that the same soft color interaction models,
using the same value for this single new parameter, give a good description of
the single diffractive hard scattering phenomena observed at the Tevatron: $W$,
dijets and beauty mesons. Also the observed rate of double leading proton
events (DPE), conventionally interpreted as double Pomeron exchange, is well
reproduced by the SCI and GAL models. Here, the same soft color interaction
mechanism produces two leading protons with associated rapidity gaps in the
same event and it is a nontrivial result that the correct rate of DPE events
are produced. 

Another, even more striking effect of two observables in the same event being
explained with the soft color interaction mechanism is diffractive $J/\psi$
production. Here, both a rapidity gap is produced and a color octet \ccbar \
pair is turned into a color singlet such that a charmonium state can be
produced. As a result we have predicted a rate of diffractive $J/\psi$
production which seems to be in good agreement with the recent preliminary CDF
result. Data on inclusive charmonium and bottomonium production (without gap
requirements) are also reproduced, as demonstrated in \cite{SCI-onium} for the
case of high-$p_\perp$ $J/\psi$, $\psi '$ and $\Upsilon$ at the Tevatron and in
\cite{Cristiano} for $J/\psi$ and $\psi '$ production at fixed target energies.

Diffractive events at the Tevatron were first obtained based on the observation
of rapidity gaps. CDF has also obtained samples defined by measured leading
antiprotons in their Roman pot detectors. Compared to the gap definition, this
gives consistent results on diffractive rates, but provides additional
information. We have used this to test details of the models, in particular the
treatment of the hadron remnant which is poorly constrained from data. Here,
one has to address issues like the treatment of a complex remnant containing
several partons and the hadronization of systems with small invariant mass. 

Comparing the different diffractive hard scattering processes we find a general
tendency that their ratio to the corresponding nondiffractive processes
decreases with increasing scale ($m_{J/\psi}$, $m_{\perp \, b}$, 
$p_{\perp\, jet}$, $m_W$, $m_Z$) of the hard process. This behavior arises
naturally in the models due to two effects. The first is the simple kinematical
correlation that an increased hard scale requires a larger momentum fraction
$x$ of the incoming parton, leaving less to the hadron remnant and thereby a
reduced probability for a leading proton with large $x_F$. The second effect is
more pQCD parton radiation which can populate rapidity regions such that no gap
is formed. This decrease of the diffractive ratio with increasing hard scale is
somewhat stronger in the GAL model than in the SCI model. This is related to the
larger cutoff for parton showers in GAL, leaving less room for radiation at
lower hard scales in particular. Furthermore, the interaction probability in
GAL depends on the invariant masses of parton pairs, making string
reconnections from high-$p_\perp$ partons more likely than from low-$p_\perp$
ones. The experimental measurements do not yet have high enough precision to
provide any clear conclusions on this scale dependence.

We have also compared the results of the Pomeron model to our models and to
data. With Pomeron parton density parametrizations obtained from diffractive
DIS at HERA, the Pomeron model gives diffractive rates at the Tevatron that are
clearly too large. The problems of the Pomeron approach have been discussed
together with possible modifications, \eg \ of the Pomeron flux, to obtain the
correct diffractive rates. There are, however, other more detailed observables
that may be used to discriminate between the models. An example was here
presented in terms of the SCI model giving different apparent branching ratios
of the $W$ in the diffractive sample. A $q\bar{q}$ from the $W$ decay will take
part in the soft color interactions and affect the probability for gap
formation, whereas leptonic $W$ decays will not have this effect. This means
that the sample of $W$ events with a gap requirement becomes biased to having
more leptonic $W$ decays. So far, diffractively produced $W$'s have only been
reconstructed through their leptonic decays. Future measurements of hadronic
$W$ decays in diffractive events are required to explore this difference of the
models regarding apparent branching ratios. We note that this effect will not be
present for diffractive $J/\psi$ or beauty mesons, since their life times are
long enough that their decay products will be produced outside the color
background field of the primary interaction. 

New data from Run II at the Tevatron with increased luminosity can give valuable
new information and higher precision diffractive data. These can provide more
decisive tests of the models and discriminate between them, perhaps ruling out
some model. In any case, additional data will constrain the models where
variations are presently possible, in particular concerning the treatment of
the hadron remnants and the formation of leading particles. Application of the
models to new processes will also be of interest. We are presently
investigating diffractive Higgs production, which will be reported in a
forthcoming paper. 

Our studies of these soft color interaction models have demonstrated that they
are able to reproduce many different phenomena: diffractive hard scattering
both in DIS at HERA and at the Tevatron as well as production of heavy
quarkonia in hadron interactions at different energies. This is quite
remarkable in view of the simplicity of the models, introducing only one new
free parameter. It also indicates that these models incorporate some essential
features of soft QCD. Therefore, the soft color interaction models
should provide guidance for the development of a proper theoretical description
of nonperturbative QCD. 

\acknowledgments We are grateful to A.\ Edin and J.\ Rathsman for many helpful
and stimulating discussions, and to A.\ Brandt, K.\ Goulianos, L.\ Motyka and
C.\ Royon for discussions and a critical reading of the manuscript. 



\begin{references}
\bibitem{StCroix} G.\ Ingelman, Diffractive hard scattering, in proc.\ {\em
Advanced Study Institute on Techniques and Concepts of High Energy Physics},
edited by T.\ Ferbel (Kluwer academic publishers, 1999), p.\ 597,
hep-ph/9912534.

\bibitem{j-g-j}
F.~Abe {\it et al.}  (CDF Collaboration),
Phys.\ Rev.\ Lett.\  {\bf 74}, 855 (1995); 
{\it ibid.} {\bf 80}, 1156 (1998);
{\it ibid.}   {\bf 81}, 5278 (1998);
B.~Abbott {\it et al.}  (D{\O} Collaboration),
Phys.\ Lett.\ B {\bf 440}, 189 (1998). 

\bibitem{singlet-exchange}
R.\ Enberg {\it et al.}, to appear in proc.~9th International 
Workshop on Deep Inelastic Scattering (DIS 2001), hep-ph/0106323, and paper in 
preparation.

\bibitem{GoulianosReview}
K.\ Goulianos, Phys.\ Rept.\ {\bf 101}, 169 (1983).

\bibitem{IS}
 G.\ Ingelman and P.E.\ Schlein, Phys.\ Lett.\ {\bf 152B}, 256 (1985). 

\bibitem{HERA-pomeron} 
T.\ Ahmed \etal (H1 collaboration), Phys.\ Lett.\ B {\bf 348}, 681 (1995);\\
M.\ Derrick \etal (ZEUS collaboration), Z.\ Phys.\ C {\bf 68}, 569 (1995).

\bibitem{HERA-F2D} 
C.~Adloff {\it et al.}  (H1 Collaboration),
Z.\ Phys.\ C {\bf 76}, 613 (1997).

\bibitem{Alvero} 
L.\ Alvero, J.C.\ Collins, J.\  Terron and J.J.\ Whitmore, 
Phys.\ Rev.\ D {\bf 59}, 074022 (1999);\\
L.\ Alvero, J.C.\ Collins and J.J.\ Whitmore, hep-ph/9806340.

\bibitem{Collins-Factorization}
J.~C.~Collins,
Phys.\ Rev.\ D {\bf 57}, 3051 (1998);
[{\it Erratum-ibid.}\ D {\bf 61}, 019902 (1998)].

\bibitem{CDF-AP}
T.\ Affolder \etal{} (CDF Collaboration), Phys.\ Rev.\ Lett.\ {\bf 84}, 5043
(2000).

\bibitem{SCI}
A.\ Edin, G.\ Ingelman and J.\ Rathsman, Phys.\ Lett.\ B {\bf 366}, 371 (1996).

\bibitem{unified}
A.\ Edin, G.\ Ingelman and J.\ Rathsman, Z.\ Phys.\ {\bf C75}, 57 (1997).

\bibitem{GAL}
J.\ Rathsman, Phys.\ Lett.\ B {\bf 452}, 364 (1999).

\bibitem{Lepto} 
G.\ Ingelman, A.\ Edin and J.\ Rathsman, Comput.\ Phys.\ Commun.\ 
{\bf 101}, 108 (1997).

\bibitem{Pythia} 
T.~Sj\"ostrand, Comput.\ Phys.\ Commun.\ {\bf 82}, 74 (1994).

\bibitem{lund}
B.\ Andersson, G.\ Gustafson, G.\ Ingelman and T.\ Sj\"ostrand,
Phys.\ Rep.\ {\bf 97}, 31 (1983).

\bibitem{CDF-W} 
F.~Abe \etal{} (CDF Collaboration), Phys.\ Rev.\ Lett.\ {\bf 78}, 2698
(1997).

\bibitem{CDF-B}
T.~Affolder \etal{} (CDF Collaboration), Phys.\ Rev.\ Lett.\ {\bf 84}, 232
(2000).

\bibitem{CDF-JJ}
F.\ Abe \etal{} (CDF Collaboration), Phys.\ Rev.\ Lett.\ {\bf 79}, 2636
(1997). 

\bibitem{D0-JJ}
B.~Abbott {\it et al.} (D0 Collaboration),
hep-ex/9912061.

\bibitem{CDF-DP}  
M.\ Albrow, FERMILAB-CONF-98-138-E (1998).

\bibitem{CDF-DPE}
T.\ Affolder \etal{} (CDF Collaboration),
Phys.\ Rev.\ Lett.\  {\bf 85}, 4215 (2000).

\bibitem{SCI-onium} 
A.\ Edin, G.\ Ingelman and J.\ Rathsman, Phys.\ Rev.\ {\bf D 56}, 7317 (1997).

\bibitem{DLpomeron} 
A.~Donnachie and P.V.~Landshoff, Phys.\ Lett.\ {\bf 191B}, 309 (1987);
[{\it Erratum-ibid.\ } B {\bf 198}, 590 (1987)].

\bibitem{fluxrenorm}
K.\ Goulianos, Phys.\ Lett.\ B {\bf 358}, 379 (1995);\\  
K.~Goulianos and J.~Montanha,
Phys.\ Rev.\ D {\bf 59}, 114017 (1999).

\bibitem{Erhan-Schlein}
S.\ Erhan and P.E.\ Schlein, Phys.\ Lett.\ B {\bf 427}, 389 (1998). 
  
\bibitem{CFL-Wproduction}
B.~E.~Cox, J.~R.~Forshaw and L.~L\"onnblad,
hep-ph/0012310.

\bibitem{Dino-new}
K.~Goulianos,
J.\ Phys.\ G {\bf 26}, 716 (2000).

\bibitem{CollinsFrankfurtStrikman}
  J.C.\ Collins, L.\ Frankfurt and M.\ Strikman, Phys.\ Lett.\ B {\bf 307}, 161
(1993).

\bibitem{Khoze}
V.~A.~Khoze, A.~D.~Martin and M.~G.~Ryskin,
Phys.\ Lett.\ B {\bf 502}, 87 (2001)
\\
A.~B.~Kaidalov, V.~A.~Khoze, A.~D.~Martin and M.~G.~Ryskin,
hep-ph/0105145.

\bibitem{Landshoff-Paris}
P.V.\ Landshoff, in proc.\ {\em Workshop on DIS and QCD}, Paris 1995, 
edited by J.F.\ Laporte, Y.\ Sirois 
(Paris, France, Ecole Polytechnique, 1995), p. 371.

\bibitem{pQCD-pomeron}
N.N.\ Nikolaev and B.G. Zakharov, Z.\ Phys.\ {\bf C53} (1992) 331;\\
M.\ W\"usthoff, Phys.\ Rev. {\bf D56}, 4311 (1997);\\
J.~Bartels, J.~Ellis, H.~Kowalski and M.~W\"usthoff,
Eur.\ Phys.\ J.\  {\bf C7}, 443 (1999);\\
M.\ W\"usthoff and A.D.\ Martin, J.Phys. {\bf G25}, R309-R344 (1999).

\bibitem{DGLAP}
V.~N.~Gribov and L.~N.~Lipatov,
Sov.\ J.\ Nucl.\ Phys.\  {\bf 15}, 438 (1972);\\ 
G.~Altarelli and G.~Parisi,
Nucl.\ Phys.\ B {\bf 126}, 298 (1977);\\
Yu.~L.~Dokshitzer,
Sov.\ Phys.\ JETP {\bf 46}, 641 (1977).

\bibitem{sce-heramc}
A.\ Edin, G.\ Ingelman and J.\ Rathsman, in proc.\ {\it Monte Carlo generators
for HERA physics}, edited by A.T.\ Doyle, G.\ Grindhammer, G.\ Ingelman and H.\
Jung, DESY-PROC-1999-02, p.\ 280, hep-ph/9912539.

\bibitem{SCI-forward}  
  A.\ Edin, G.\ Ingelman and J.\ Rathsman, in proc. {\it Future physics at
  HERA}, edited by G.\ Ingelman, A.\ De Roeck and A.\ Klanner, DESY 96-235, p.\
  580.

\bibitem{leading-pn}
C.~Adloff {\it et al.}  (H1 Collaboration),
Eur.\ Phys.\ J.\  {\bf C6}, 587 (1999).

\bibitem{Multiple} 
T.\ Sj\"ostrand and M.\  van Zijl, Phys.\ Lett.\ {\bf 188B}, 149 (1987);
Phys.\ Rev.\ {\bf D36}, 2019 (1987).

\bibitem{Underlying} 
G.~J.~Alner {\it et al.}  (UA5 Collaboration),
Phys.\ Lett.\  B {\bf 138}, 304 (1984);\\
G.~Arnison {\it et al.}  (UA1 Collaboration),
Phys.\ Lett.\  {\bf B132}, 214 (1983).

\bibitem{Pythia6} 
T.~Sj\"ostrand {\it et al.}, Comput.\ Phys.\ Commun.\ {\bf 135}, 238 (2001).

\bibitem{CTEQ3} 
H.L.~Lai \etal{} (CTEQ Collaboration), Phys.\ Rev.\ {\bf D51}, 4763 (1995).

\bibitem{CTEQ4}
H.L.~Lai \etal{} (CTEQ Collaboration), Phys.\ Rev.\ {\bf D55}, 1280 (1997).

\bibitem{Pompyt} 
P.~Bruni and G.~Ingelman, DESY-93-187 in proc.\ International Europhysics
Conference on High Energy Physics, Marseille 1993, p. 595; \\
P.~Bruni, A.~Edin and G.~Ingelman, {\tt http://www3.tsl.uu.se/thep/pompyt/}.

\bibitem{GSpomeron} 
T.~Gehrmann and W.J.~Stirling, Z.\ Phys.\ {\bf C70}, 89 (1996).

\bibitem{Bruni+I}
P.\ Bruni and G.\ Ingelman, Phys.\ Lett. B {\bf 311}, 317 (1993). 

\bibitem{UA8-1}
R.~Bonino {\it et al.}  (UA8 Collaboration),
Phys.\ Lett.\ B {\bf 211}, 239 (1988).

\bibitem{UA8-2}
A.~Brandt {\it et al.}  (UA8 Collaboration),
Phys.\ Lett.\ B {\bf 297}, 417 (1992);
Phys.\ Lett.\ B {\bf 421}, 395 (1998).

\bibitem{D0-DP}
D{\O} collaboration, Hard Diffractive Jet Production at D{\O}, presented at
{\it XXIX International Conference on High Energy Physics, ICHEP 98}, July
1998, Vancouver, Canada;\\
K.\ Mauritz, Nucl.\ Phys.\  (Proc.\ Suppl.)  {\bf B79}, 378 (1999).

\bibitem{Cristiano} 
C.~B.~Mariotto, M.~B.~Gay Ducati and G.~Ingelman,
Hard and soft QCD in charmonium production, 
hep-ph/0008200;\\ 
M.~B.~Gay Ducati, G.~Ingelman and C.~B.~Mariotto, 
Soft and Hard QCD Dynamics in $J/\psi$ Hadroproduction, 
Uppsala report TSL/ISV-2001-0241 and paper in preparation.
                 
\bibitem{johan-quarkonium}
J.~Rathsman,
in proc.\ {\it Monte Carlo generators for HERA physics}, edited by A.T.\
Doyle, G.\ Grindhammer, G.\ Ingelman and H.\ Jung, DESY-PROC-1999-02, p.\ 421,
SLAC-PUB-8343, TSL-ISV-99-0219. 

\bibitem{Goulianos}
K.\ Goulianos, FERMILAB-CONF-99-154-E (1999). 

\bibitem{Nachtmann}
O.~Nachtmann and A.~Reiter,
Z.\ Phys.\ C {\bf 24}, 283 (1984);\\
G.~W.~Botz, P.~Haberl and O.~Nachtmann,
Z.\ Phys.\ C {\bf 67}, 143 (1995).

\bibitem{Buchmueller}
W.~Buchm\"uller and A.~Hebecker,
Nucl.\ Phys.\ B {\bf 476}, 203 (1996);
see also the review in
A.~Hebecker,
Phys.\ Rept.\  {\bf 331}, 1 (2000).


\end{references}
\end{document}